\documentclass[nojss]{jss}
\usepackage{amssymb}
\usepackage{amsmath}
\usepackage{rotating}

\author{Christian Panse\\Swiss Federal Institute of Technology Zurich }

\title{Rectangular Statistical Cartograms in \proglang{R}: The \pkg{recmap} Package}

\Plainauthor{Christian Panse} 
\Plaintitle{Compute the Rectangular Statistical Cartogram} 
\Shorttitle{\pkg{recmap}: Compute the Rectangular Statistical Cartogram} 

\Abstract{
Cartogram drawing is a technique for showing geography-related 
statistical information, such as demographic and epidemiological data. 
The idea is to distort a map by resizing its regions according to a 
statistical parameter by keeping the map recognizable. 
This article describes an R package implementing an algorithm called 
RecMap which approximates every map region by a rectangle  
where the area corresponds to the given statistical value (maintain zero 
cartographic error).
The package implements the computationally intensive tasks in 
\proglang{C++}.
This paper's contribution is that it demonstrates on real and synthetic 
maps how \pkg{recmap} 
can be used, how it is implemented and used with other statistical packages.
}
\Plainkeywords{cartogram, spatial data analysis, geovisualization, demographics, R} 

\Submitdate{2016-06-01}

\Address{
  Christian Panse\\
  Functional Genomics Center Zurich UZH\texttt{|}ETHZ\\
  Winterthurerstr. 190\\
  CH-8057, Z{\"u}rich, Switzerland\\
  Telephone: +41/44/63-53912
  E-mail: \email{cp@fgcz.ethz.ch}\\
  URL: \url{http://www.fgcz.ch/the-center/people/panse.html}
}

\begin{document}


\newcommand{\abs}[1]{\vert#1\vert}

\section{Introduction}
The idea to generate a cartogram is to distort a map by
resizing its regions according to a given statistical
parameter, but in a way that keeps the map recognizable.
These so-called cartograms or {\em value-by-area maps}
may be used to visualize any geo-spatial related data,
e.g., political, economical, or public health data. There
exists several algorithms to compute 
so-called contiguous
cartograms. An overview on historic, hand-drawn, and
computer generated cartograms
can be found in \citet{Waldo, TheStateoftheArtInCartograms}.

For using {\em contiguous cartograms}, the diffusion-based method
of \citet{pmid15136719} is available through
the R packages \pkg{Rcartogram} and \pkg{getcartr}
\citet{Rcartogram,getcartr}.

An alternative approach to {\em contiguous cartograms} is to entirely relax the
map topology by approximating each map region by basic geometric objects like
rectangles or circles \citep{circle}.
Such rectangular cartograms can be generated from geolocation and statistical data. Hence, they provide a useful alternative, even if there are no boundaries available or some statistical values are missing.
First rectangular cartograms were drawn by hand following a system of construction
\citep{ErwinRaisz}. Recent research publication on rectangular cartogram drawing include
\cite{Speckmann2004,Speckmann2007,Speckmann2012,Buchin:2016}.
However, according to a recent publication, both variants of
\pkg{recmap} are the only rectangular cartogram algorithms that ``maintain zero
cartographic error'' \citep[section 5.4]{TheStateoftheArtInCartograms}.

\begin{figure}[htbp]
\centering
\includegraphics[width=\textwidth,keepaspectratio]{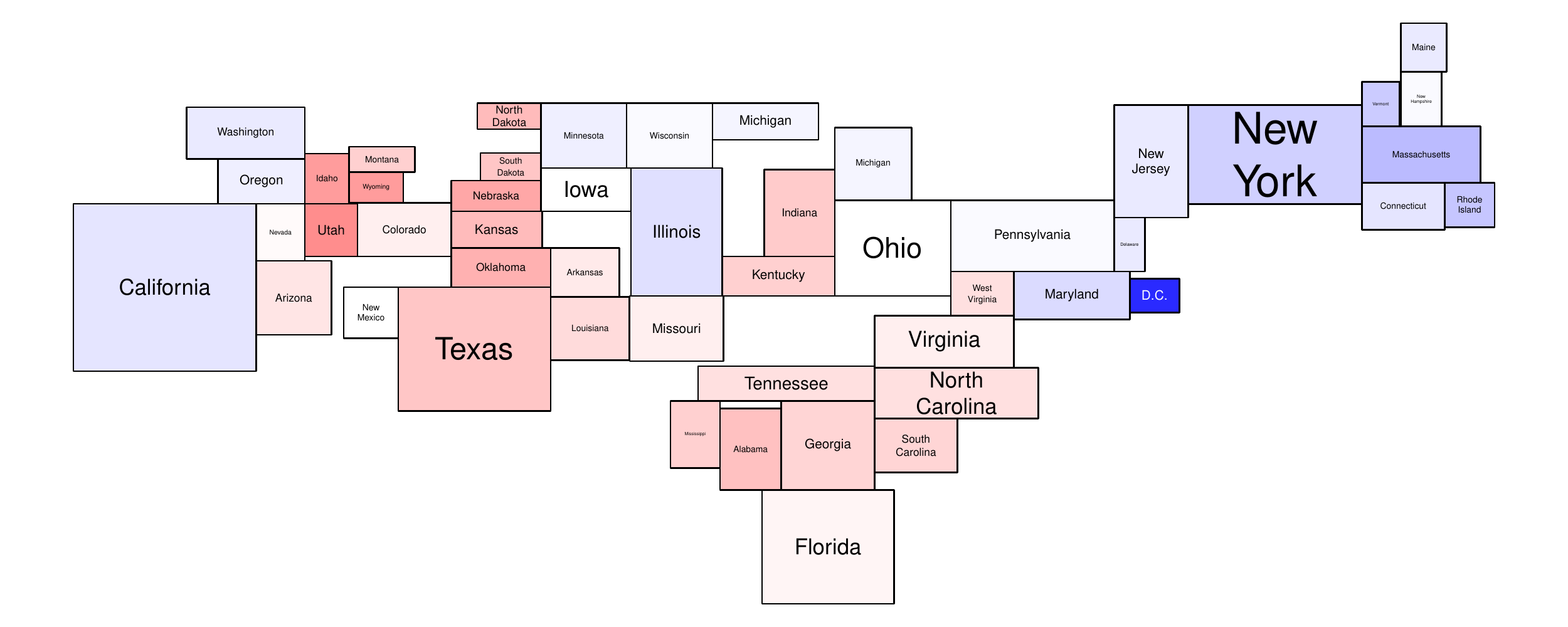}
\caption{A rectangular statistical cartogram of the U.S. election in 2004 is
drawn. The area corresponds to the number of electors. Color is indicating
the outcome of the vote.
Democrats are represented by the color blue and republicans are
represented through red coloring.
Regions with low saturation, e.g., Ohio, Pennsylvania, and Florida,  mark states
with a tight outcome of the vote (also known as swing states).
The election cartogram was computed by
using the original implementation of the construction heuristic
RecMap MP2 introduced by \citet{recmap}.
map source: U.S. Census Bureau;
election data source: \url{http://www.electoral-vote.com/}, November 2004.}
\label{figure:USelection}
\end{figure}

The \pkg{recmap} package discussed in this article
contains an implementation of the RecMap (Map Partition
variant 2) algorithm \cite{recmap} to draw maps
according to given statistical parameter. A typical usage
of a cartogram based visualization is demonstrated in
Figure~\ref{figure:USelection}.

The article is organized as follows: The next section
defines the input and output of the \code{recmap} algorithm and
the objective functions. In Section~\ref{section:usage},
the usage of the \pkg{recmap} package using the R shell is
demonstrated. Section \ref{section:implementation}
discusses major implementation details and provides some
benchmark and performance studies. Section~\ref{section:metaheuristic} describes how two
metaheuristics can be used to find an optimized cartogram
drawing. In Section~\ref{section:application}, a number
of applications are presented.
Section~\ref{section:summary} summarizes and discuss the approach.

\newpage
\section{Problem definition and objective functions}
\label{section:obj}

\paragraph{The input} consists of a map represented by overlapping
rectangles $\mathcal{R} = (r_1, \ldots, r_n)$.
Each map region $r_j$ contains:

\begin{itemize}

\item a tuple of $(x, y)$ values corresponding to the
(longitude, latitude) position,

\item a tuple of $(dx, dy)$ of expansion along
(longitude, latitude),

\item and a statistical value $z$.

\end{itemize}

The $(x, y)$ coordinates represent the center of the
minimal bounding boxes (MBB).
The coordinates of the MBB
are derived by adding or subtracting the $(dx, dy)$ tuple
accordingly. A tuple $(dx, dy)$ defines also the ratio
of the corresponding map region. The statistical values 
$(z_1, \hdots, z_n)$ define the
desired area of each map region.

The ordering $\Pi$ is an index vector taken from the permutation set $Perm(n)$.

\paragraph{The output} is a rectangular cartogram
$\mathcal{\overline{R}}$ where the map regions are:

\begin{itemize}

\item non overlapping,

\item connected,

\item rectangles are placed parallel to the axes.

\end{itemize}

Furthermore, for each map region
the following criteria have to be satisfied:

\begin{itemize}

\item the area is equal to the desired area derived from the
as input given statistical value $z$,

\item the ratio, $dy/dx$, is preserved.

\end{itemize}

The \code{recmap} construction heuristic is a function
\begin{eqnarray}
\label{eqn:f}
f:\mathbb{R}^{n \times 2} \times \mathbb{R}_{>0}^{n \times 3} \times Perm(n)	&\rightarrow&	\mathbb{R}^{n \times 2} \times \mathbb{R}_{>0}^{n \times 2}
\end{eqnarray}
which transforms the set of input rectangles $\mathcal{R}$
and a permutation $\Pi$
into a rectangular cartogram
\begin{eqnarray}
\label{eqn:Rout}
\overline{\mathcal{R}}&=&f(\mathcal{R}, \Pi)
\end{eqnarray}
so that important spatial constraints,
in particular

\begin{itemize}
\item the topology of the dual graph $G(\mathcal{R}, E)$,
defined by the overlapping input rectangles,
\item the relative position of map region centers,
\end{itemize}
are tried to be preserved.

If the output satifies these criteria, the rectangular
cartogram is denoted as {\em feasible solution}.

The following equation were introduced by
\citet[Definition 1]{cartodraw}. The desired area
$\tilde{A_j}$ of a map region $r_j$ is defined as

\begin{eqnarray} 
\label{eqn:Adesired}
\tilde{A_j}&=& z_j \cdot \frac{\sum_{i=1}^{n}A(r_i)} {\sum_{i=1}^{n} z_{i}}
\end{eqnarray}

where the area of the rectangle $r$ is defined by
\begin{eqnarray}
\label{eqn:area}
A(r) &=& 4 \cdot dx \cdot dy
\end{eqnarray}
The objective functions for area $d_A$, shape $d_S$ (ratios of the MBBs),
relative positon $d_R$, and map topology
$d_T$, are as defined and described by \citet[equations 2--4]{recmap}:

\begin{eqnarray}
\label{eqn:obj0}
d_A&=&d_A(\mathcal{R,\overline{R}})\\
&=&\sum_{j=1}^{n} \abs{A_j - \tilde{A_j}} \\
d_S&=&d_S(\mathcal{\mathcal{R},\overline{R}})\\
\label{eqn:dS}
&=&\sum_{j=1}^{n} \abs{ (dy_j/dx_j) - (\overline{dy_j}/\overline{dx_j}) }\\
d_T&=&d_T(\mathcal{\mathcal{R},\overline{R}})\\
\label{eqn:dT}
&=&\frac{|\overline{E}_a\backslash E_a|+ |E_a\backslash\overline{E}_a|}{|\overline{E}_a\cup E_a|},\label{form_topology}\\
d_R&=&d_R(\mathcal{R},\mathcal{\overline{R}})\\
\label{eqn:dR}
\label{eqn:objn}
&=&\frac{2}{n\cdot\left(n-1\right)}\cdot\sum\limits_{i=1}^{n-1}\sum\limits_{j=i+1}^{n} \abs{\measuredangle (\underbrace{r_i, r_j}_{\in \mathcal{R}}) - \measuredangle(\underbrace{\tilde{r_i}, \tilde{r_j}}_{\in \mathcal{\overline{R}}})}
\end{eqnarray}

Note, if $\forall j \in \{1, \hdots, n\}: A_j = \tilde{A_j} \Rightarrow d_A = 0$
and if $\forall j \in \{1, \hdots, n\}: dy_j/dx_j = \overline{dy_j}/\overline{dx_j} \Rightarrow d_S=0$.
$n=\abs{\mathcal{R}}$ and $E$ denotes the edges of the dual graph.
$dT$ determines the differences in the pseudo dual graphs. $dR$ measures the angle differences between all pairs of rectangle centers of the input map and output cartogram by using  $\measuredangle (x, y)$ defined as follows\footnote{Implemented using the \proglang{C++} method \code{std::atan2} -- ``Computes the arc tangent of y/x using the signs of arguments to determine the correct quadrant.'' \url{http://en.cppreference.com/w/cpp/numeric/math/atan2}, access: 2017-01-01.}:
\begin{eqnarray}
\label{eqn:atan2}
\measuredangle (x, y) &=& \arctan_2(x, y)
\end{eqnarray}

To find an optimal rectangular statistical cartogram out of all feasible solutions,
as it will be intended in this manuscript using \pkg{recmap},
the optimization problem can be expressed as follows:
\begin{eqnarray}
& \underset{\Pi \in Perm(n)}{\text{minimize}}  &  a \cdot d_R + b \cdot d_T \text{~with~} a, b \in \mathbb{R}_{\geq 0} \\
& \text{subject to}  & d_A = 0, d_S = 0.
\end{eqnarray}

To find a sufficiently good solution for the optimization problem
a metaheuristic, as described in section \ref{section:metaheuristic},
will be applied.

\begin{table}
\centering
\begin{tabular}{llr}
\hline
\hline
Term  & Description & Equation\\
\hline
$\mathcal{R} = (r_1, \ldots, r_n)$  & overlapping input rectangles  & \\
$(x, y)$  & coordinates represent the center of a rectangle  & \\
$(dx, dy)$  & expansion along x and y axes  & \\
$z$ & statistical value of a rectangle  & \\
$G(\mathcal{R}, E)$ & dual graph of $\mathcal{R}$ & \\
$\Pi$ & permutation & \\
$A(r)$  & area of a rectangle  & \label{eqn:area}\\ 
$\tilde{A_j}$ & desired area of a map region $j$ & \ref{eqn:Adesired}\\
\hline
$d_A$ & area error  & \ref{eqn:obj0}\\
$d_S$ & shape error & \ref{eqn:dS}\\
$d_T$ & topology error & \ref{eqn:dT}\\
$d_R$ & relative position error & \ref{eqn:dR}\\
$\measuredangle (x, y)$ & angle between two points $x$ and $y$ in $\mathbb{R}^2$  & \ref{eqn:atan2}\\
\hline
$f$ & construction heuristic  & \ref{eqn:f}\\
$\overline{\mathcal{R}}$  & output / cartogram  & \ref{eqn:Rout}\\
\hline
\hline
\end{tabular}
\caption{
The table provides a glossary of the used algebraic terms.
\label{table:glossary}
}
\end{table}
\section{The package usage}
\label{section:usage}
\subsection{Input}

The U.S. map on state level is often used to compare
cartogram algorithms. To generate a useful dual graph,
which is a requirement of the algorithm, the
input map regions have to overlap. 
For the \code{state.x77} data used in this section
this can be done by
correcting lines of longitude derived by the square roots
of the area values (see Figure~\ref{figure:usage} left).

\begin{Schunk}
\begin{Sinput}
R> US.map <- data.frame(x=state.center$x,
+      y = state.center$y,
+      dx = sqrt(state.area) / 2 / (0.7 * 60 * cos(state.center$y * pi / 180)),
+      dy = sqrt(state.area) / 2 / (0.7 * 60) ,
+      z = sqrt(state.area),
+      name = state.name)
R> head(US.map)
\end{Sinput}
\begin{Soutput}
          x       y        dx       dy        z       name
1  -86.7509 32.5901  3.209890 2.704478 227.1761    Alabama
2 -127.2500 49.2500 14.005670 9.142338 767.9564     Alaska
3 -111.6250 34.2192  4.859044 4.017906 337.5041    Arizona
4  -92.2992 34.7336  3.338204 2.743370 230.4431   Arkansas
5 -119.7730 36.5341  5.902177 4.742415 398.3629 California
6 -105.5130 38.6777  4.923602 3.843727 322.8730   Colorado
\end{Soutput}
\end{Schunk}

Please note, in general \code{recmap} is not transforming the
geodetic datum, e.g., WGS84 or Swissgrid.
Furthermore, geospatial positions have to be mapped from the
earth surface to the plane.
This has to be done prior the the cartogram generation.
An overview of map
projection can be found in \citet{snyder1997flattening}.
Map projections aim to optimize towards different objectives, e.g., 
conformal mapping (preservation of local angles) or area mapping (preservation of area).
In cartogram publications the map projection aspect is
often neglected and conformal errors are accepted.
However, studying U.S. cartograms,
a cylindrical projection seems to be the projection of choice.
The \proglang{R} user can find support for a wide variety
of adequate map projections due to using the \pkg{mapproj}
package by \cite{mapproj}.

\subsection{Run}

The algorithm takes a \code{data.frame} object having the column 
names \code{c('x', 'y', 'dx', 'dy', 'z', 'name')}, here the \code{US.map}
object, as input. Additional control is given by the index order
and the (dx, dy) values. The index order $\Pi$ defines in which
order the dual graph is explored and the (dx, dy) values
define which map region are adjacent.

\begin{Schunk}
\begin{Sinput}
R> library("recmap")
R> # Generate the rectangular statistical cartogram.
R> US.cartogram <- recmap(US.map)
\end{Sinput}
\end{Schunk}

\subsection{The output}

The \code{recmap} method returns an S3 class object \code{c('recmap', 'data.frame')} of the
transformed map. Additional columns in the result contain
information for topology and relative position error defined in Equations 
\ref{eqn:dT} and \ref{eqn:dR}.
The column \code{dfn.num} indicates in which order the dual graph has been 
explored.

\begin{Schunk}
\begin{Sinput}
R> head(US.cartogram)
\end{Sinput}
\begin{Soutput}
           x        y       dx       dy        z       name dfs.num
1  -79.27226 10.40598 3.916894 3.300161 227.1761    Alabama      23
2 -129.22283 63.71115 8.181797 5.340748 767.9564     Alaska      49
3 -126.86923 30.51915 4.819169 3.984933 337.5041    Arizona      34
4  -87.19357 10.42302 3.994415 3.282650 230.4431   Arkansas      39
5 -137.00972 33.40013 5.311325 4.267664 398.3629 California      35
6 -115.35010 62.23315 4.851078 3.787109 322.8730   Colorado      48
  topology.error relpos.error relposnh.error
1              6    0.3917837      0.7848231
2              6    0.5901459      1.9059785
3              3    0.2447048      0.3498989
4              8    0.5624989      1.0728192
5              4    0.1861821      0.2596710
6             10    0.7576109      1.4491548
\end{Soutput}
\end{Schunk}

\begin{figure}[htbp]
\centering
\resizebox{1.0\textwidth}{!}{
	\includegraphics{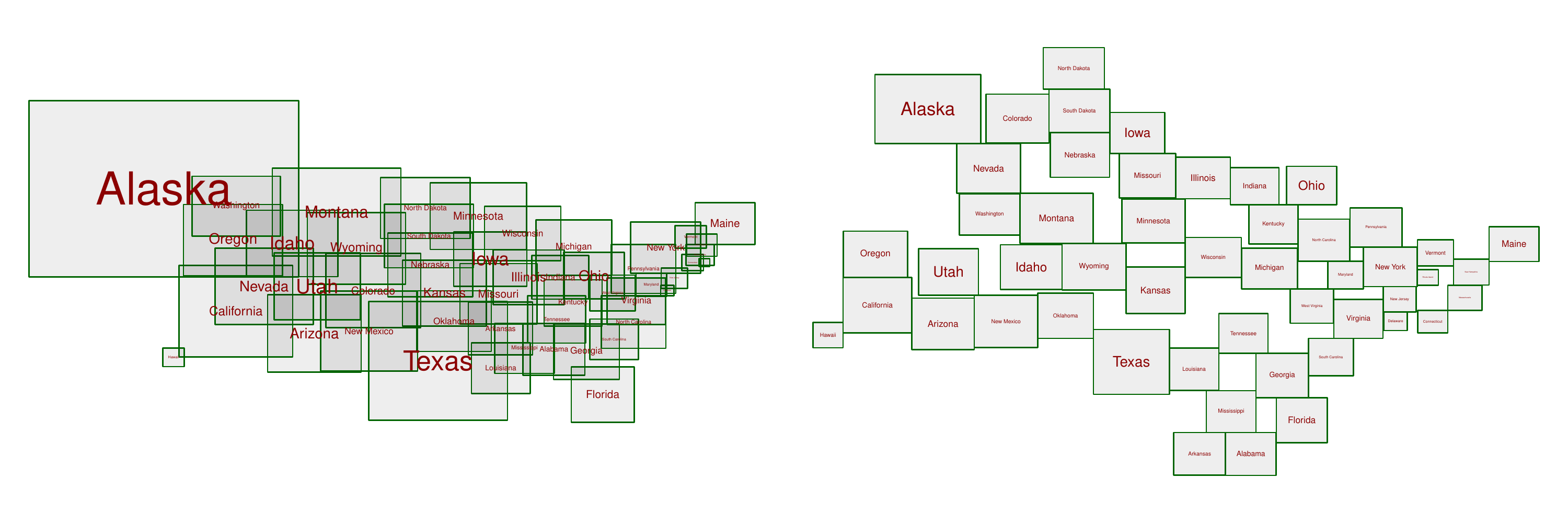}
}
\caption{The ``usage'' example, generated from the ``US State Facts and Figures''
\code{datasets} package, was used for drawing a rectangular
map approximation.
The input set of overlapping rectangles is shown left.
A feasible solution thereof generated with
\code{recmap} is on the right side.\label{figure:usage}
The state area, original size of the map region, is used as statistical value.}
\end{figure}

Input and output of the \code{recmap} function can be
visualized using the S3 method \code{plot.recmap}. The output of the R
code snippet can be seen in Figure~\ref{figure:usage}.
As default the \code{plot.recmap} function places the name attribute in
the centre of each rectangle. To avoid overplotting of the labels,
the text is scaled by using \code{cex = dx / strwidth(name)}
as argument for the \code{text} function.
However small statistical values result in small label areas.
This problem can be  circumvented by using an interactive visualization,
e.g., using \pkg{shiny} the function \code{hoverOpts} enables
the ``mouseover'' feature.
Please note: while the input \code{US.map} is not classified as a \code{recmap} class, 
the \code{plot.recmap} function can not be dispatched by the S3 class system
and has to be explicitly called.

\begin{Schunk}
\begin{Sinput}
R> op <- par(mfrow = c(1, 2), mar = c(0, 0, 0, 0))
R> plot.recmap(US.map, col.text = 'darkred')
R> plot(US.cartogram, col.text = 'darkred')
\end{Sinput}
\end{Schunk}

A summary method implements the calculation of some meta data including the objective functions.
\begin{Schunk}
\begin{Sinput}
R> summary(US.cartogram)
\end{Sinput}
\begin{Soutput}
                             values
number of map regions     50.000000
area error                 0.000000
topology error           266.000000
relative position error    0.420000
screen filling [in %]     36.893585
xmin                    -146.967409
xmax                     -34.722706
ymin                       7.105818
ymax                      73.190904
\end{Soutput}
\end{Schunk}

\newpage

\section{Implementation}
\label{section:implementation}

\begin{figure}[htbp]
  \centering
\includegraphics[width=0.49\textwidth]{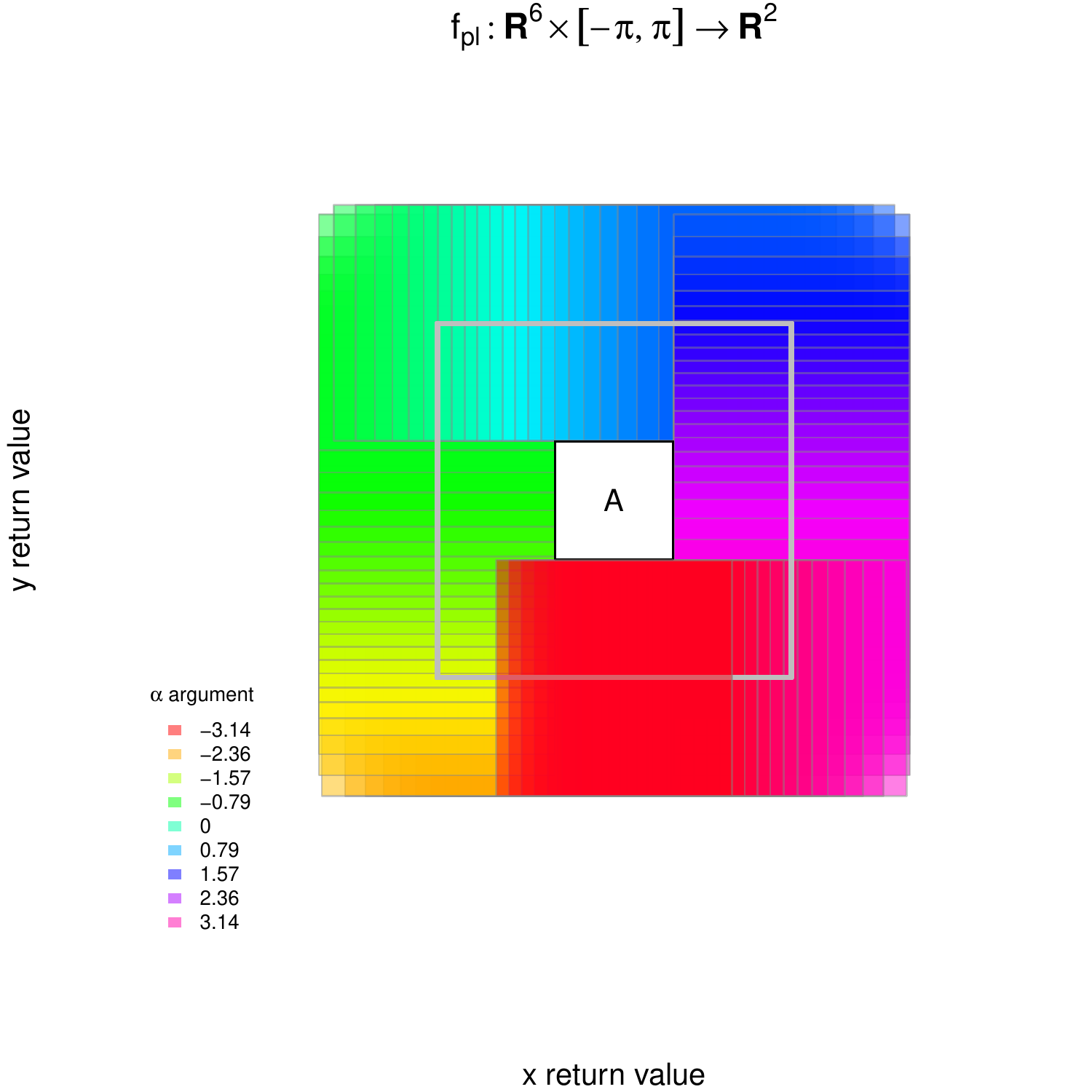}
\includegraphics[width=0.49\textwidth]{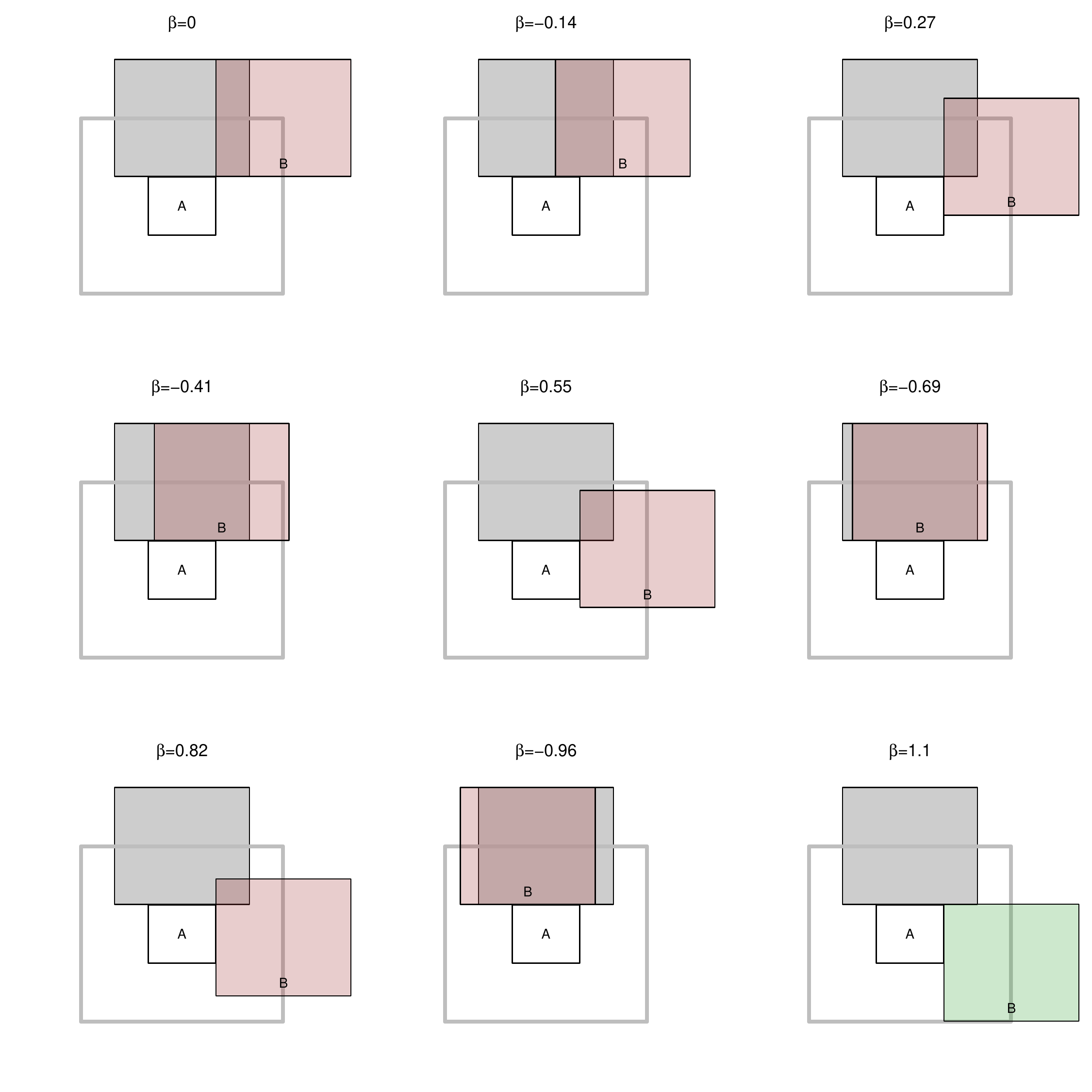}
\caption{The rainbow colored rectangles graph the feasable positions of 
a local placement function $f_{pl}$ to place a rectangle $B$ 
around a given rectangle $A$
of any angle $\alpha$ between $[-\pi,\pi]$ (left figure).
There are special cases for quadrant I, II, III, and IV indicated by different
colors in the graphic.
On the right figure, B should be placed around A starting with an angle of
$\frac{\pi}{4}$ (this is the initial relative position in the input map). 
Since the rectangle 
can not be placed without overlap (indicated by red) an angle $\beta$ is added.
\label{figure:bearing}}
\end{figure}

\pkg{recmap} is implemented in \proglang{C++} using features provided by
the \proglang{C++-11} standard. The input and output data transfer between R and
the \proglang{C++} \code{recmap} class is handled by using the \pkg{Rcpp} \citep{Rcpp} mechanism.
The \proglang{C++} \code{recmap} class itself consists of a \code{std::vector} of
\code{map\_region}. A \code{map\_region} contains all the $(x, y, dx, dy, z)$
values, a \code{std::vector} of type \code{int} to it neighbor map regions,
and some additional help variables to ease the error computation. In general
the construction algorithm follows the map partition 2 (MP2) procedure
described in \cite{recmap}. 
The local placement function can place any rectangle next to another rectangle 
as it is demonstrated in Figure \ref{figure:bearing}.
The current implementation starts with the original bearing $\alpha$ of the two map
region centers (see Figure \ref{figure:bearing} where $\beta=0$). 
If the placement does not lead to a feasible solution,
the angle $\beta$ is added to $\alpha$.
$\beta$ is iterating between $[0, \pi]$ with step size of $\frac{\pi}{180}$ and a
changing sign until a placement without overlap has been found. 
If no placement can be found, the algorithm considers all
adjacent placed map regions.
If also in a later step during the DFS a map region can not be placed,
a non feasible
solution is accepted. This situation is often caused because the construction
algorithm is hampered by the input
configuration of the map regions. Solving this can be very compute expensive
and often the procedure leads to a solution which will be rejected by the
metaheuristic due to the bad fitness value.

Furthermore, no genetic algorithm (metaheuristic) has been implemented.
Here \pkg{recmap} will use the \pkg{GA} package available on CRAN
as demonstrated in section \ref{section:metaheuristic}. The most computationally
expensive part is the computation of MBB intersections which has to be
performed to achieve feasible solutions, multiple times, for each placement
step. In the package version 0.2.1 these tests were performed by iterating
over each map region. All later versions use a \code{std::multiset} data
structure and a \code{std::lower\_bound} algorithm 
of the \proglang{C++} Standard Template Library (STL)
to reduce the search
space.

\begin{figure}[htbp]
  \centering
\includegraphics[width=1.0\textwidth]{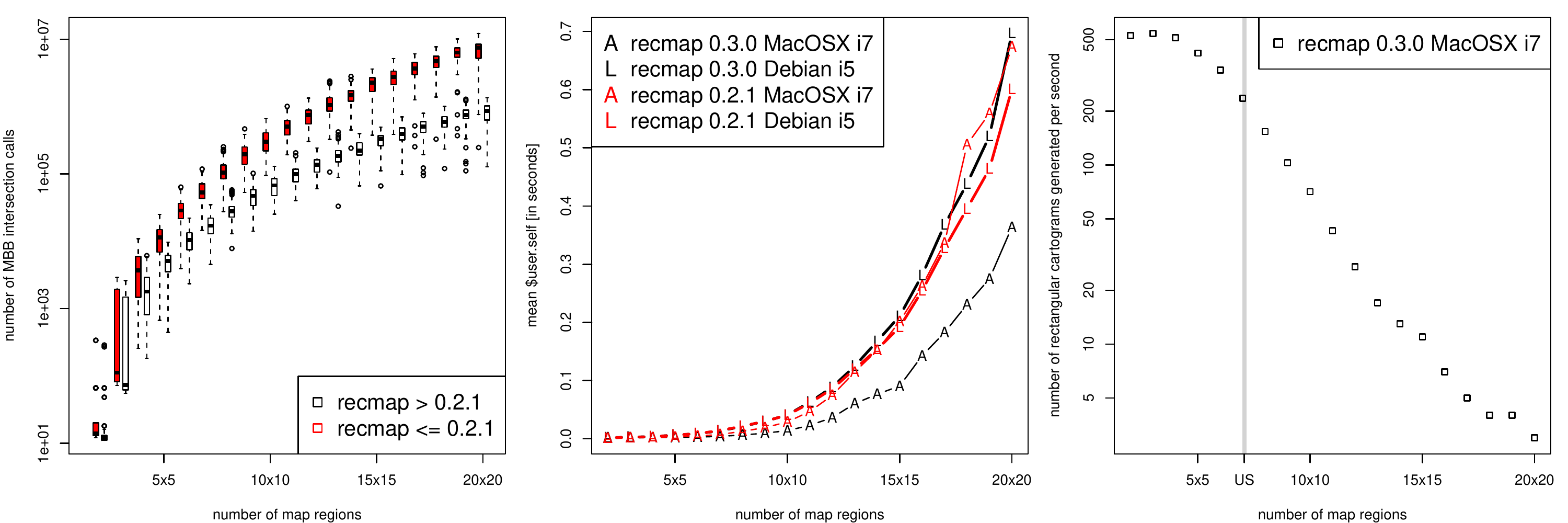}
\caption{The boxplot (left) graphs the number of MBB intersection test calls for a
given checkerboard size. The graph in the middle displays the mean aggregated
measured computation time.
The scatterplot (right) displays how many rectangular cartograms can be generated
within one second depending on the board size.
\label{figure:benchmark}}
\end{figure}

The time complexity for one \code{recmap} run is $\mathcal{O}(n^2)$, where $n$ is the number of
regions. A depth first search (DFS) run is visiting each map region only
once and therefore it has time complexity $\mathcal{O}(n)$. For each
\code{map\_region} placement a constant number of MBB intersection are
called (max 360). MBB check is implemented using \code{std::multiset}
container, and the functions \code{std::insert}, \code{std::upper\_bound},
and \code{std::upper\_bound}. The time complexity for all of these functions
are reported on \url{http://www.cplusplus.com/reference/stl} as
$\mathcal{O}(\log(n))$. However, the worst case scenario for a range query is
$\mathcal{O}(n)$, iff dx or dy cover the whole x or y range. The boxplots in
Figure~\ref{figure:benchmark} (left plot) show that the number of MBB
intersection test calls could be reduced by using a \code{std::multiset} data
structure. For this benchmark, synthetic
checkerboards with a number of map regions in the interval of $\left[2^2, \ldots,
20^2\right]$ were generated using the R function
\code{checkerboard}. For each checkerboard size \code{recmap}
was called 100 times using different index orderings of the checkerboard
input. The index order of the input records have a direct impact how the
DFS is traversing the map. This characteristic will later be used for the
metaheuristic.

For a performance study the checkerboard bench described above was repeated
using the \pkg{rbenchmark} package by \cite{rbenchmark} to measure the \code{proc.time}. The
study was performed on two systems (using one core only). An Intel(R) Core(TM)
i5-2500 CPU @ 3.30GHz running a Debian 8 using a 3.16.0-4-amd64 Kernel and a
3 GHz Intel Core i7 (Apple MacBook Pro) running OS X 10.11.4 (15E65) using a
Darwin 15.4.0 Kernel. The middle graphic in Figure~\ref{figure:benchmark}
displays the resulting mean aggregated measured data of the two (hardware /
OS) systems using the two different implementations of the MBB intersection
test. Beside the fact that the Linux system can not benefit from the more
efficient implementation of the MBB intersection test,
the graphic (middle) shows even for an input
size of $20^2=400$ map regions, a rectangular cartogram can be computed in
less than a second.

The right plot in Figure~\ref{figure:benchmark} was derived from the
performance study (plot in the middle). It shows the number of rectangular
cartograms which can be generated within one second depending on the number
of map regions. The grey vertical line indicates the number of the U.S. map
regions. The ability to generate a high number of cartogram candidates in a
short time period is an important requirement for any metaheuristic as it is
demonstrated in the next section.

\section{Choose a metaheuristic}
\label{section:metaheuristic}

The design of the \pkg{recmap} algorithm is as follows: First, compute a set
of feasible solutions. In a second evaluation step choose the best solution.
In the current implementation, variations can be introduced
by changing the index order $\Pi$ of the input data.
This has a direct impact to the DFS traversal
and leads to different cartogram layouts.
The objective functions, equations \ref{eqn:obj0} to \ref{eqn:objn},
can be used to evaluate the result and to define a
fitness function which has to be maximized.
The following R code defines the
fitness function which will be used as default.

\begin{Schunk}
\begin{Sinput}
R> recmap:::.recmap.fitness
\end{Sinput}
\begin{Soutput}
function (idxOrder, Map, ...) 
{
    Cartogram <- recmap(Map[idxOrder, ])
    if (sum(Cartogram$topology.error == 100) > 0) {
        return(0)
    }
    1/sum(Cartogram$relpos.error)
}
<environment: namespace:recmap>
\end{Soutput}
\end{Schunk}

Other variants of fitness functions will lead to different results as it is shown
in \cite[Figure 4]{recmap}.

Since it is not possible to compute and evaluate all permutations,
which is $n!$,
random experiments are conducted.
In the following section it is demonstrated how two metaheuristic,
GRASP and GA, can be used
to find a optimal layout for the rectangular statistical cartogram.

For a visual evaluation, of the in this section proposed metaheuristics,
an 8x8 checkerboard will be used as input map.
The map is generated using the \code{checkerboard} method
\begin{Schunk}
\begin{Sinput}
R> Checkerboard <- checkerboard(8)
R> summary(Checkerboard)
\end{Sinput}
\begin{Soutput}
                        values
number of map regions    64.00
area error                0.27
topology error              NA
relative position error     NA
screen filling [in %]   100.00
xmin                      0.50
xmax                      8.50
ymin                      0.50
ymax                      8.50
\end{Soutput}
\end{Schunk}
and can be seen in Figure~\ref{figure:cmp_GA_GRASP} (left).
If the assumption is made, that for each vertex,
the cyclic order of edges in the contiguous cartogram 
remains the same as as in the input map,
checkerboards provide examples
of sets of map regions which do not have ideal cartogram
solutions \citep[Definition 2, Lemma 1, Fig. 3.]{cartodraw}.

\subsection{Greedy Randomized Adaptive Search Procedures}

One group of optimizer is ``trivial to efficiently implement'' \citep{GRASP} and
can directly benefit from a parallel environment is called {\em greedy
randomized adaptive search procedures} (GRASP). 
The R method \code{recmapGRASP} defines a generic GRASP implementation as described in~\cite[Fig.~1]{GRASP}.
The \code{recmapGRASP} function generates a set of rectangular
cartograms which have different layouts caused by the
random sampling. Each cartogram is evaluated. The best
candidate is saved. The following command will generate cartogram solution
based on a GRASP metaheuristic.

\begin{Schunk}
\begin{Sinput}
R> set.seed(1)
R> res.GRASP <- recmapGRASP(Checkerboard)
R> plot(res.GRASP$Cartogram,
+    col = c('white', 'white', 'white', 'black')[res.GRASP$Cartogram$z])
\end{Sinput}
\end{Schunk}

A drawing of the cartogram can be found in Figure~\ref{figure:cmp_GA_GRASP}
(middle).

For some types of input maps, GRASP can generate amazing results in a short time.
As it can be seen in Figure~\ref{figure:cmp_GA_GRASP}, for the checkerboard map,
GRASP is outperformed by the genetic algorithm, introduced in the next section.
The plot in Figure~\ref{figure:cmp_GA_GRASP2} (right) shows that the solution
process runs too fast into a saturation.

\subsection{Lessons learnt from biological evolution}

In this paragraph a constraint-based genetic algorithm
(GA) as discussed in \cite{recmap} is used as
metaheuristic. Here the construction heuristic benefits
from the existence of the \pkg{GA} package by \cite{GA}.
The GA configuration used for \code{recmap} was directly
derived from the {\em traveling salesperson problem} (TSP) example
\cite[section 4.8]{GA} using the {\em permutation} type of the \code{ga} method.
As genotype the index order $\Pi$ of the input map is used. The
following command generates an almost perfect rectangular
cartogram for the checkerboard map having 64 map regions
on the author's laptop\footnote{Mac Book Pro from 2015, please find the hardware 
specification in Table \ref{table:overview}.} 
within 60 seconds.

\begin{Schunk}
\begin{Sinput}
R> recmap.GA <- ga(type = "permutation",
+      fitness = recmap:::.recmap.fitness,
+      Map = Checkerboard,
+      min = 1, max = nrow(Checkerboard),
+      popSize = 64,
+      maxiter = 1000,
+      maxFitness = 1.7,
+      parallel = TRUE,
+      pmutation = 0.25)
\end{Sinput}
\end{Schunk}

\begin{figure}[htbp]
\centering
\includegraphics[width=1.0\textwidth,keepaspectratio]{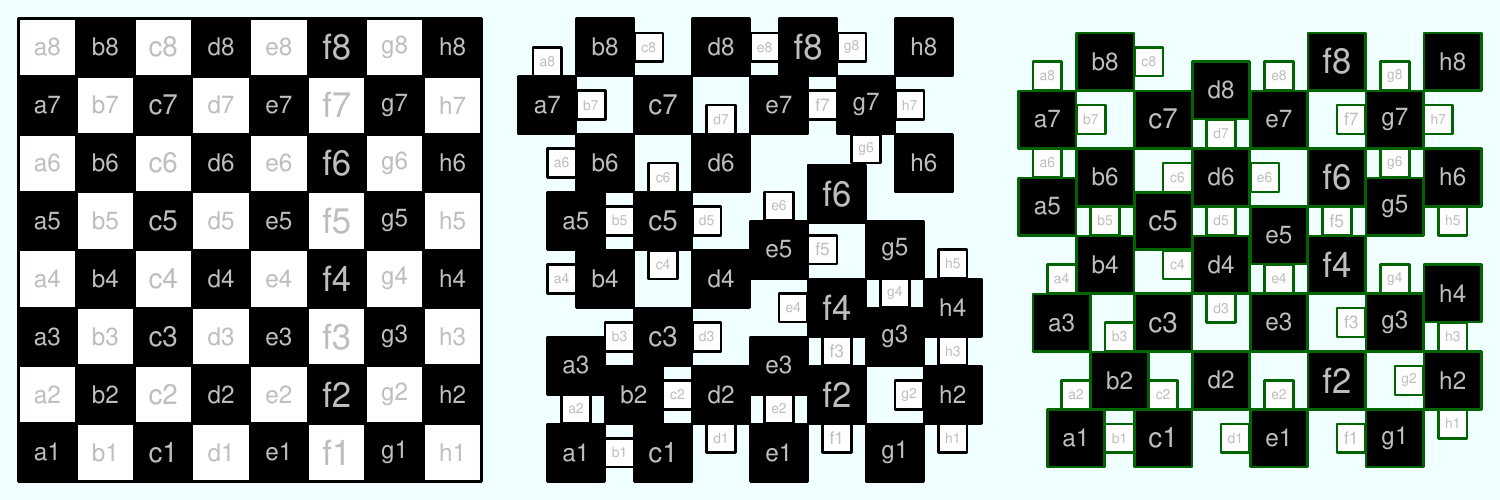}
\caption{A comparison of the input map, GRASP, and GA is
displayed. As input map a 8x8 checkerboard (left) has been generated.
The area of a black box needs to be four times as large as the area of a white
box. The cartogram in the middle proposes a solutions generated by a GRASP
metaheuristic. The right cartogram graphs a solution computed by a genetic
algorithm within one minute.}
\label{figure:cmp_GA_GRASP}
\end{figure}

The metaheuristic stops when a maximum number of
iteration has been performed or the fitness value is
higher than 1.7. Having reached a fitness value of 1.7
using the \code{recmap.fitness} function the result
in Figure~\ref{figure:cmp_GA_GRASP} (right)
looks like an almost ``optimal'' solution.

The \code{recmapGA} function is a higher level wrapper function to glue
the \code{recmap} construction heuristic with the metaheuristic \code{ga}.

\begin{Schunk}
\begin{Sinput}
R> res.GA <- recmapGA(Checkerboard)
R> summary(res.GA)
R> plot(res.GA$Cartogram, 
+    col=c('white', 'white', 'white', 'black')[res.GA$Cartogram$z])
\end{Sinput}
\end{Schunk}

\begin{figure}[htbp]
\centering
\includegraphics[height=0.49\textwidth,keepaspectratio]{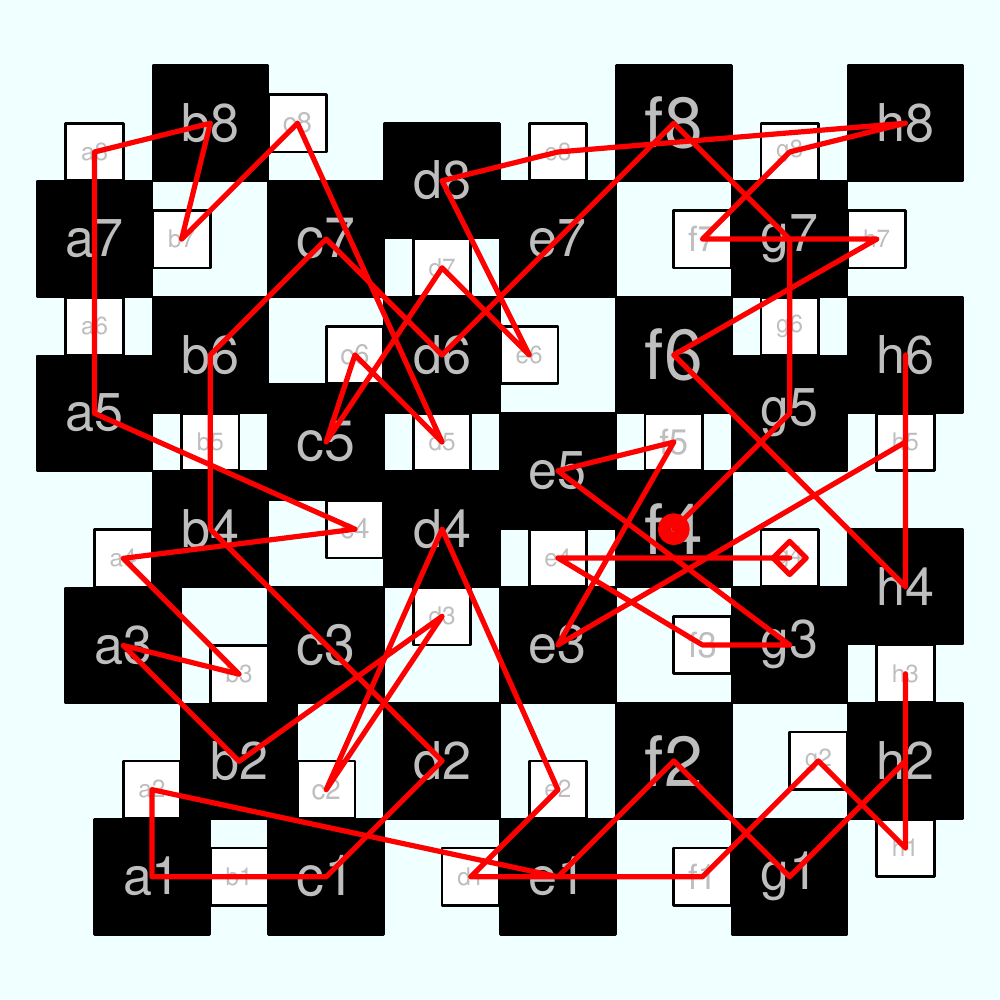}
\includegraphics[height=0.49\textwidth,keepaspectratio]{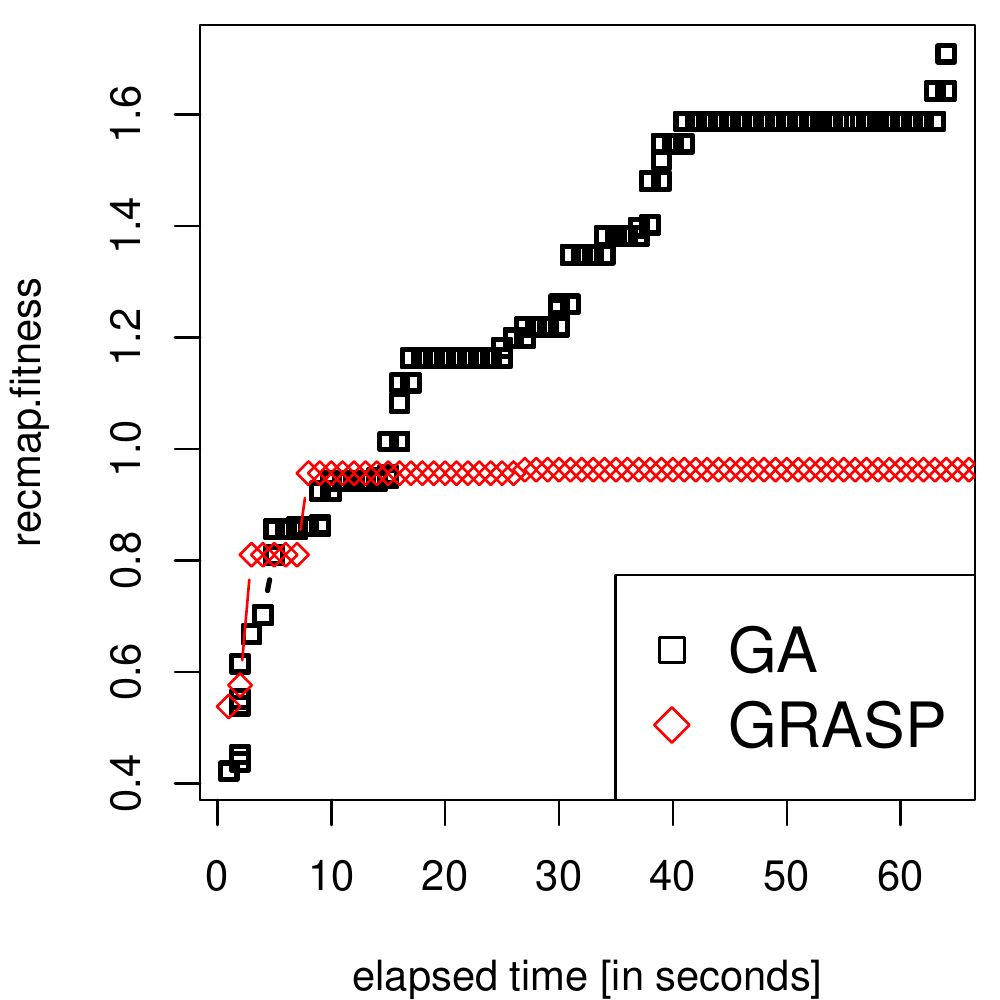}
\caption{The left map graphs the order of the placement during the cartogram
generation. The right plot displays the \code{recmap.fitness} values versus
elapsed time of the two metaheuristics (right).\label{figure:cmp_GA_GRASP2}}
\end{figure}

The \code{recmapGA} function returns a list of the input \code{Map},
The solution of the GA, and a \code{recmap} object containing the cartogram.
The resulting cartograms using a GA can be seen in
Figure~\ref{figure:cmp_GA_GRASP} (right).
The red line in Figure~\ref{figure:cmp_GA_GRASP2} (left)
indicates in which order the rectangles
were placed using the DFS numbering. The red $\bullet$ symbol marks the first
placed rectangle and the $\diamond$ the last one.

\begin{figure}[htbp]
\centering
\includegraphics[width=0.9\textwidth,keepaspectratio]{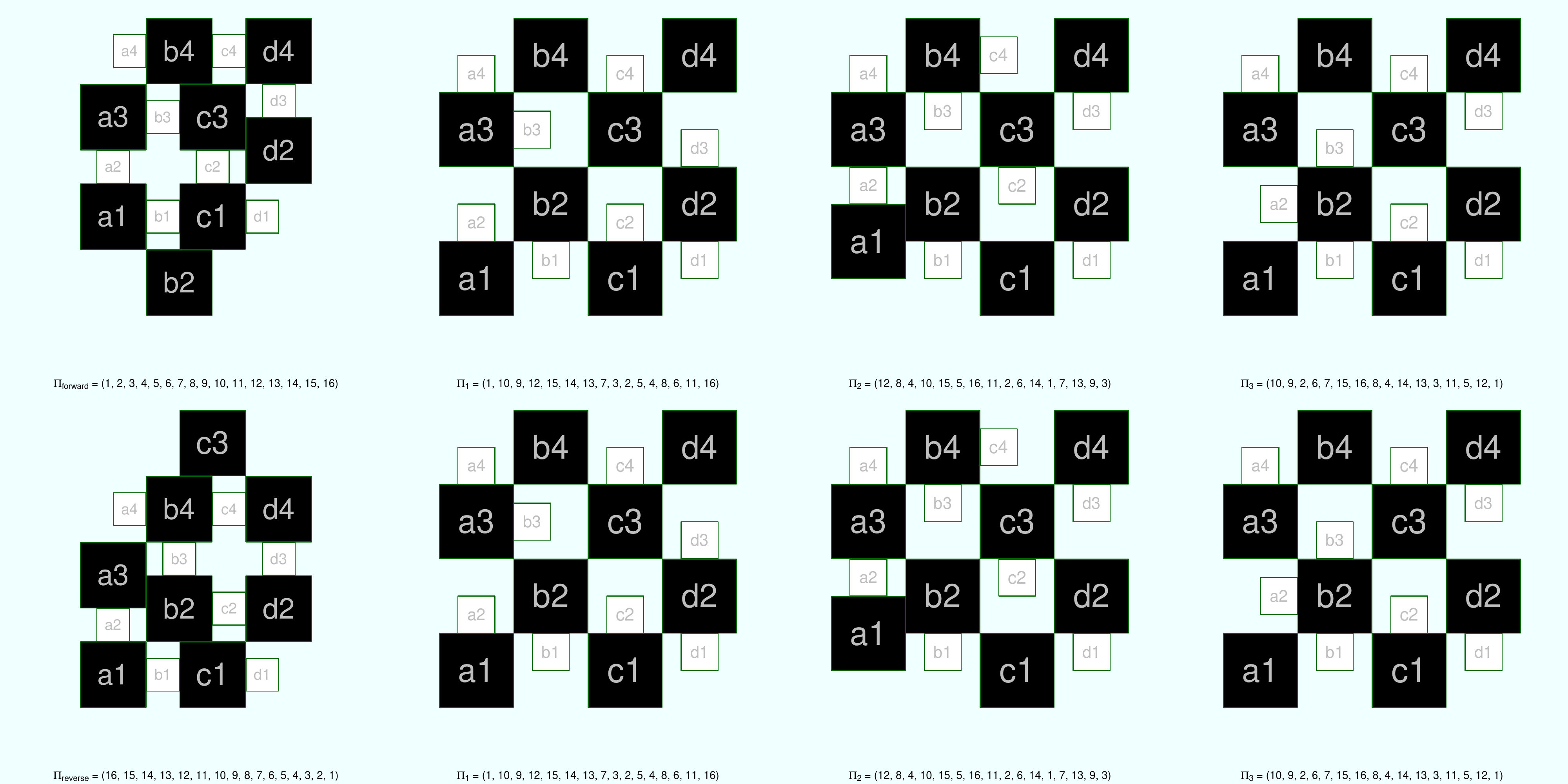}
\caption{This graphical example illustrates the variability in
solutions, dependent on the initial seed value in $\Pi_{\texttt{seed}}$.
The left column displays forward and reverse index orders. All other columns
show the results from duplicate seeds $\{1,2,3\}$ to demonstrate that the same 
seed will lead to the same index order and to the identical layout of the cartogram.
\label{figure:GAseed}}
\end{figure}

Figure \ref{figure:GAseed} illustrates the variability in solutions,
dependent on the initial seed value. The experiment was repeated twice to
demonstrate the effect that the same seed values lead to the same
permutation order $\Pi$ and finally to the same cartogram construction.

\begin{Schunk}
\begin{Sinput}
R> fitness.weighted <- function (idxOrder, Map, ...) 
+  {
+      Cartogram <- recmap(Map[idxOrder, ])
+      if (sum(Cartogram$topology.error == 100) > 0) {
+          return(0)
+      }
+      
+      S <- summary(Cartogram)
+      dT <- max(Cartogram$topology.error)
+      dR <- S[4,]
+      dE <- (100 - S[5,]) / 100
+      
+      # weighting the objectives
+      1 / (c(0.2, 0.6, 0.2) %*% c(dT, dR, dE))
+  }
\end{Sinput}
\end{Schunk}

\begin{Schunk}
\begin{Sinput}
R> set.seed(2)
R> US.map.best <- recmapGA(Map = US.map,
+           fitness = fitness.weighted, 
+           maxiter = 100,
+           maxFitness = 100,
+           popSize = 50,
+           keepBest = TRUE, 
+           pmutation = 0.35, 
+           parallel = TRUE)
\end{Sinput}
\end{Schunk}

\begin{figure}[htbp]
\centering
\includegraphics[width=0.85\textwidth,keepaspectratio]{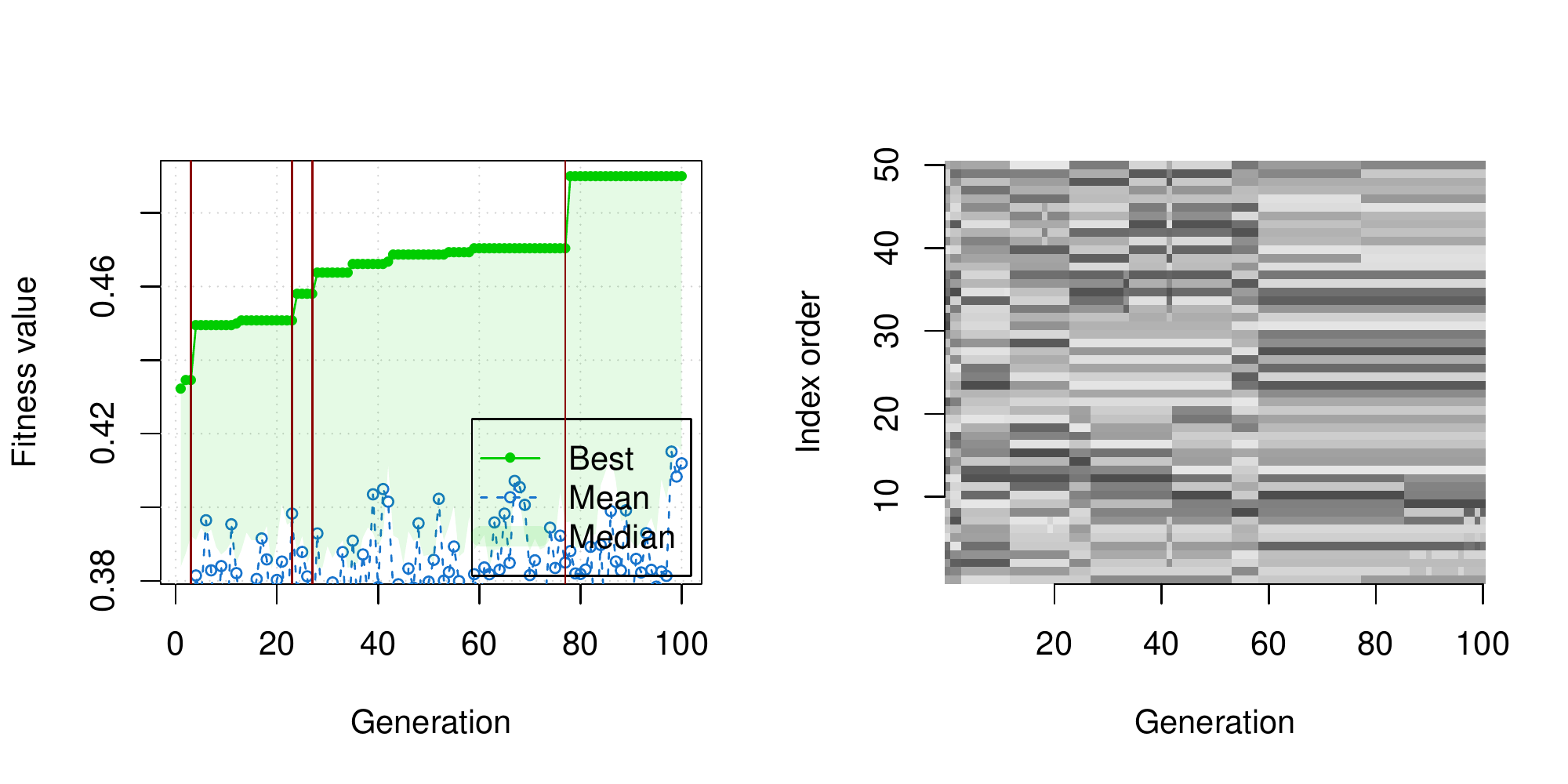}
\includegraphics[width=0.85\textwidth,keepaspectratio]{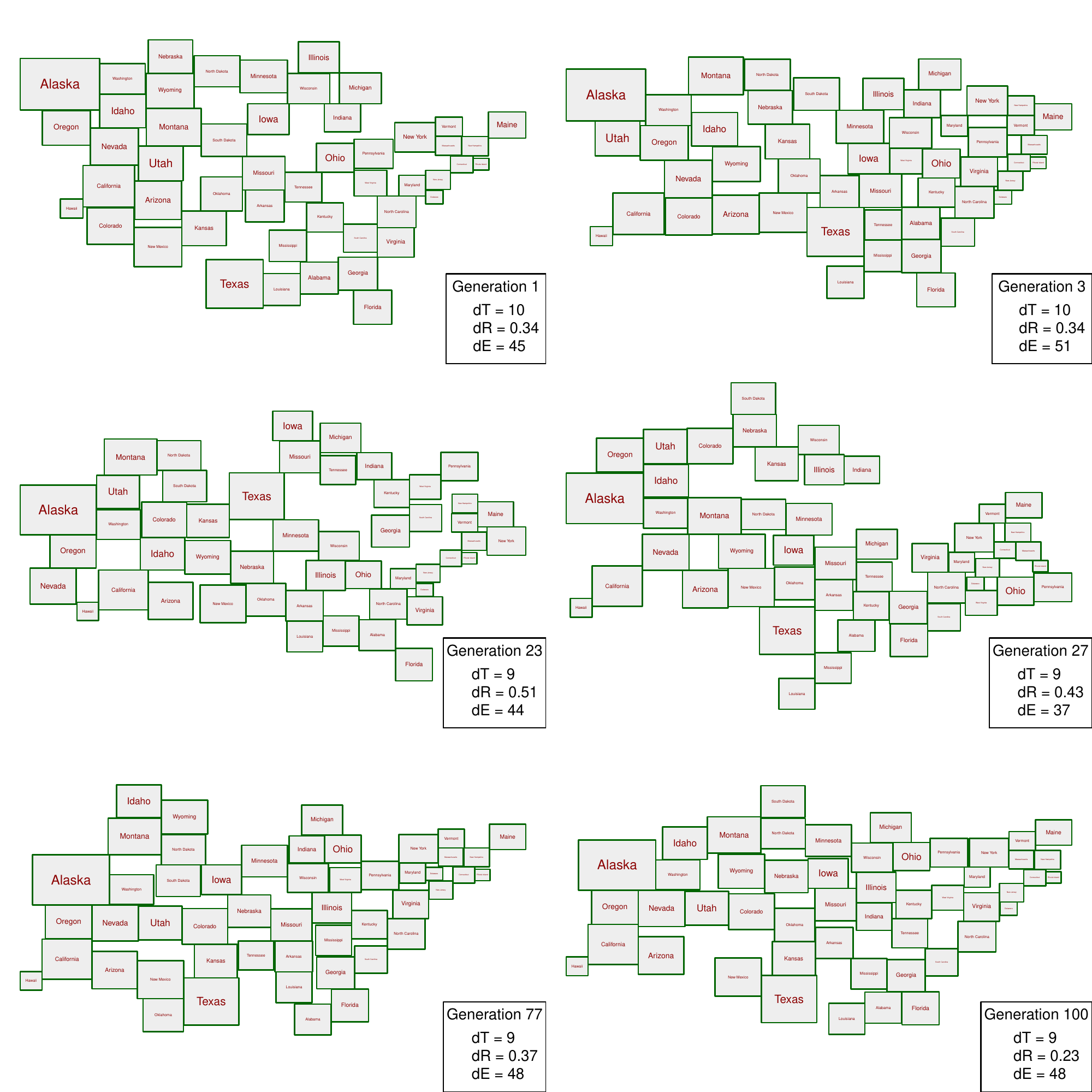}
\caption{The plot on the top left displays the fitness value during the 
increasing generation of the genetic algorithm. 
The image plot on the top right visualizes the genotype ($\Pi$) change from one
generation to the other
using a grey colormap for encoding the index order.
The six phenotypes visualize the improvement of the solution with an increasing 
number of generations.\label{figure:bestSolution}}
\end{figure}
The regtangular map approximations in Figure~\ref{figure:bestSolution}
demonstrate the continuous improvement of the feasible solutions
with an increasing number of generations using the GA as metaheuristic for the
data used in Section~\ref{section:usage}.

Note that the metaheuristic of the space filling {\em Quad Tree} \citep{Finkel1974} based 
RecMap MP1 variant could also be realized 
by using the \pkg{GA} package.
Here, instead of a permutation, a binary representation of decision variables has to be chosen.
This can be done by setting the \code{type} attribute of the \code{ga} function to \code{'binary'}.
The genotyp, given as a binary vector, is defining the split type of the
{\em Quad Tree} data structure. A \code{1} is applying a vertical split while a \code{0} 
triggers a horizontal split. 

\section{Application}
\label{section:application}

\begin{figure}[htbp]
\centering
\includegraphics[height=0.9\textheight,keepaspectratio]{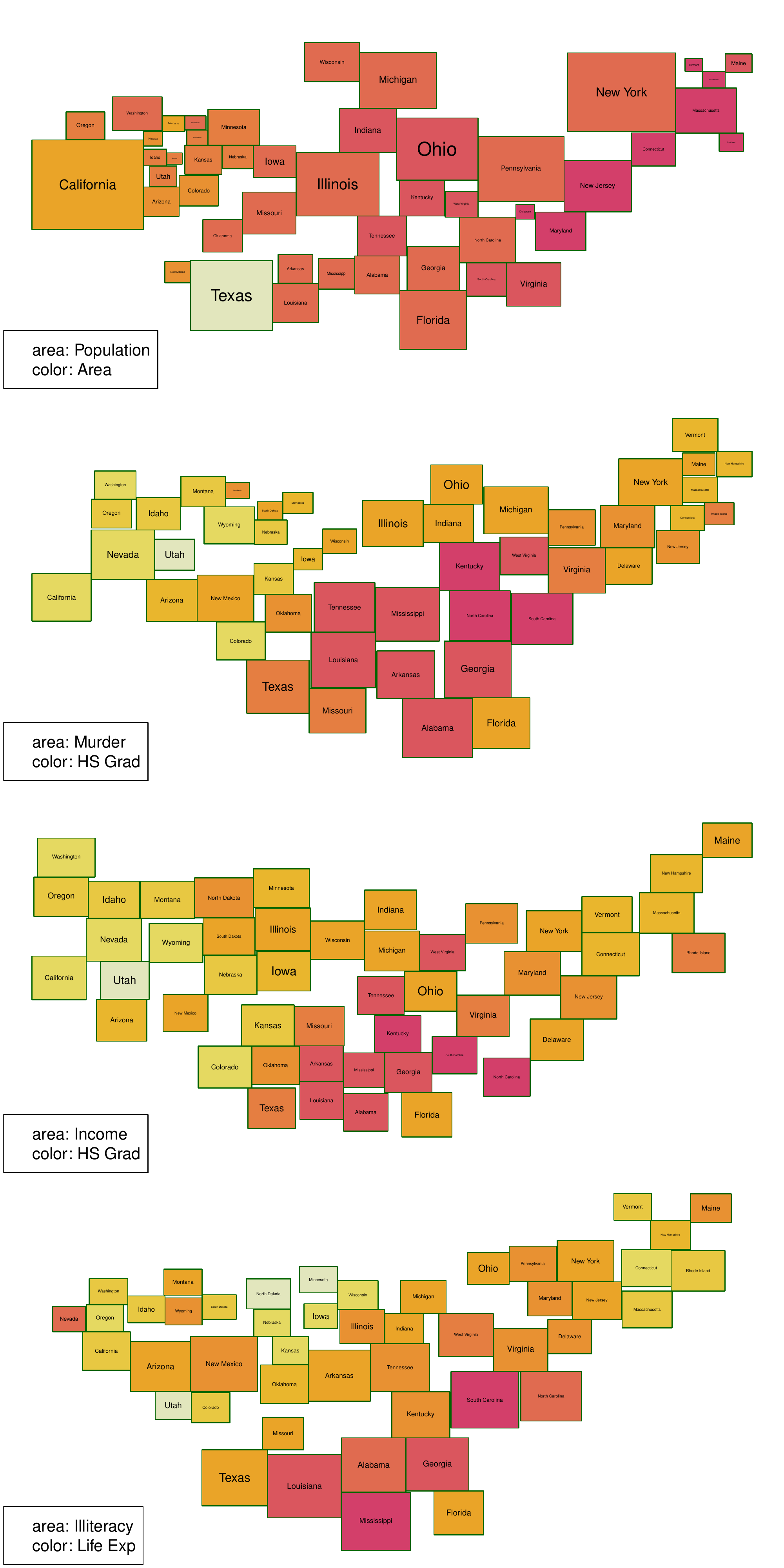}
\includegraphics[height=0.9\textheight,keepaspectratio]{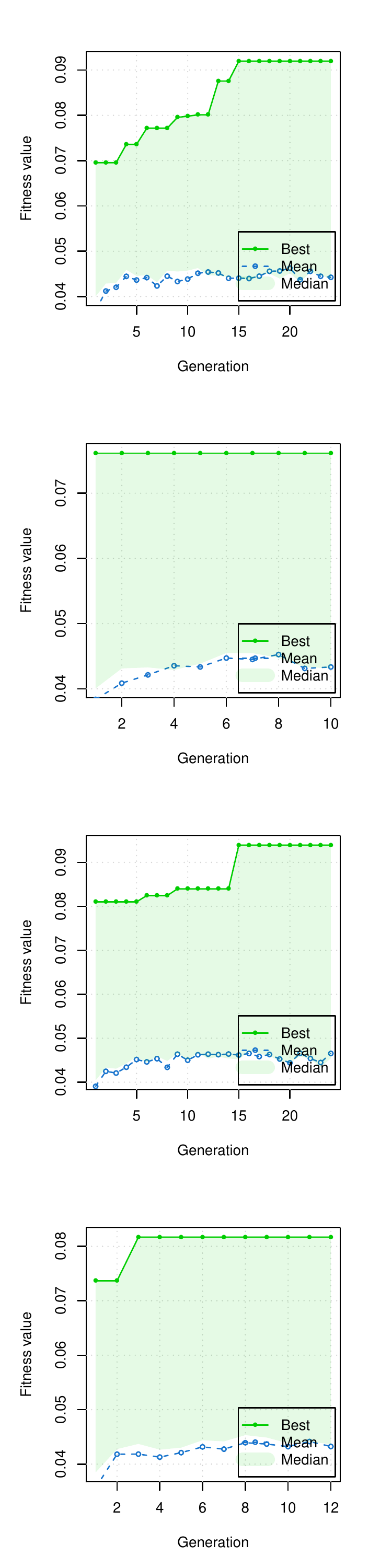}
\caption{Rectangular Statistical Cartograms using the ``US State Facts
and Figures'' dataset are drawn (see also
\url{https://recmap.shinyapps.io/state_x77/}).
The plots on the right column display the fitness values versus the generation of 
the genetic algorithm during the optimization process.
}
\label{figure:x77}
\end{figure}

This section applies \pkg{recmap} to some real world maps
having numbers of map regions of different magnitudes
and different
kind of topology. Beside Figures~\ref{figure:USelection}
and \ref{figure:x77} the examples focus more on the
demonstration of the drawing characteristics of the
\code{recmap} method itself and less on the information
visualization scopes.
Figure~\ref{fig:brexit} demonstrates how \pkg{recmap}
objects can be transformed to \code{SpatialPolygonsDataFrame}
objects.

\paragraph{US State Facts and Figures based cartograms}
are displayed in Figure~\ref{figure:x77}. The data are
available from the data frame \code{state.x77}. On the
cartograms two statistical data are displayed using the
area and the color of a map region.
The colormap was generated by using the \code{heat\_hcl}
function of the \pkg{colorspace} package by \cite{colorspace} (red is low; white is high).
The code below is a wrapper function for the
\code{recmap} and the \code{ga} functions. A tuple of
\code{state.x77} column names is given as input.

\begin{Schunk}
\begin{Sinput}
R> library("colorspace")
R> recmap_state_x77 <- 
+  function(input, Map = US.map, DF = state.x77, cm = heat_hcl(10)){
+    # join
+    Map <- cbind(Map, DF, match(Map$name, row.names(DF)))
+  
+    # filter 
+    Map <- Map[!Map$name %in% c("Hawaii", "Alaska"), ]
+  
+    # set attribute for desired area
+    Map$z <- Map[, input$area]
+  
+    ptm <- proc.time()
+    res <- recmapGA(Map = Map,
+      popSize = 300, maxiter = 30, run = 10, parallel = TRUE)
+  
+    # set attribute for the coloring 
+    S <- Map[res$GA@solution[1, ], input$color]
+    col.idx <- round((length(cm) -1)  * (S - min(S)) / (max(S)  - min(S))) + 1
+  		    
+    # have fun
+    plot(res$Cartogram, col = cm[col.idx], col.text='black')
+    legend("bottomleft", c(paste("area:", input$area), 
+                           paste("color:", input$color)), cex=1.5)
+  
+    res$time.elapsed = (proc.time()-ptm)[3]
+    res
+  }
\end{Sinput}
\end{Schunk}

As input map the \code{US.map} defined in Section~\ref{section:usage} is used.
An interactive \pkg{shiny}~\citep{shiny} web application,
using this method, provides more combination of
attributes and is available through
\url{https://recmap.shinyapps.io/state_x77/}.

\begin{Schunk}
\begin{Sinput}
R> op <- par(mfrow = c(4, 1), mar = rep(0.25, 4), bg = "white")
R> set.seed(1)
R> cartogram.x77 <- lapply(list(list(color = "Area", area = "Population"), 
+      list(color = "HS Grad", area = "Murder"), list(color = "HS Grad", 
+          area = "Income"), list(color = "Life Exp", area = "Illiteracy")), 
+      recmap_state_x77)
R> par(op)
\end{Sinput}
\end{Schunk}

\begin{Schunk}
\begin{Sinput}
R> op <- par(mar = c(5, 5, 3, 3), mfrow = c(4, 1))
R> res <- lapply(cartogram.x77, function(x) {
+      plot(x$GA)
+  })
R> par(op)
\end{Sinput}
\end{Schunk}
    
\begin{figure}[htbp]
\centering
\includegraphics[height=0.8\textheight, keepaspectratio]{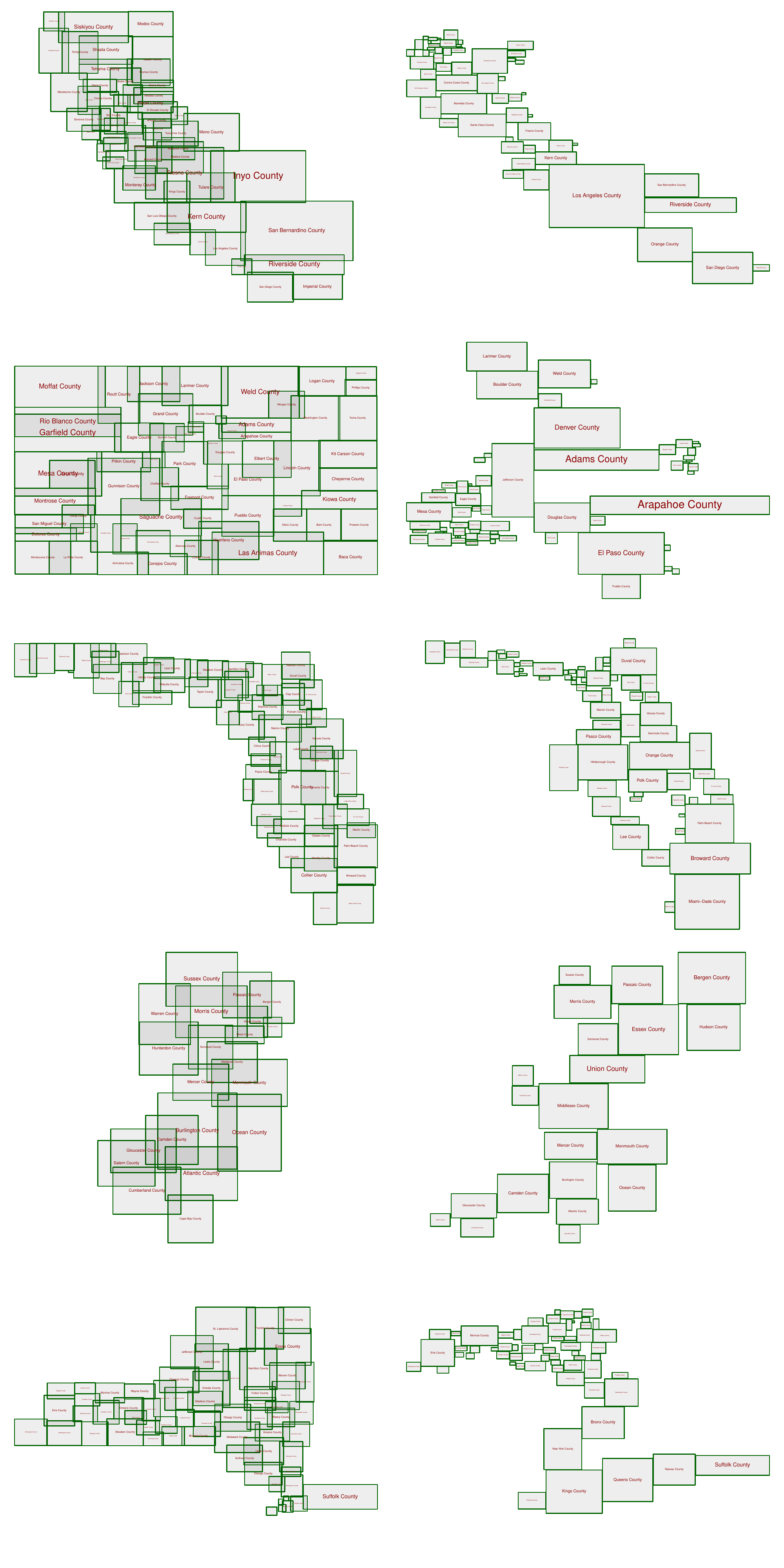}
\includegraphics[height=0.8\textheight, keepaspectratio]{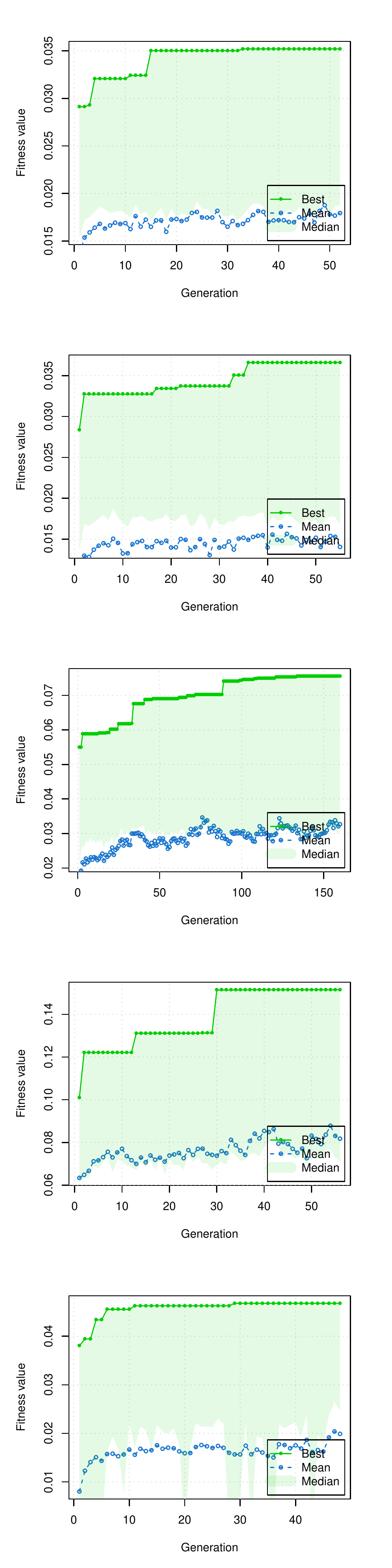}
\caption{U.S. input maps (left) of California, Colorado, Florida,
New~Jersey, and New~York on county level were used as input to compute
2010 census population cartograms (middle; top to bottom).
On the right column the fitness values versus the generations are displayed.
}
\label{figure:us_census}
\end{figure}

\paragraph{US population cartograms on county level} 
showing cartograms of
California, Colorado, Florida, New~Jersey, and New~York
are displayed in Figure~\ref{figure:us_census}. The map
material was extracted from the \pkg{maps}
package by \cite{maps}
and the population data were retrieved from the
\pkg{noncensus} package by \cite{noncensus}. The map
regions were joined over the fips (Federal Information
Processing Standard) county codes using the
\code{counties} data frame.
The cartograms were generated by using the genetic algorithm as metaheuristic.

\begin{figure}[htbp]
\centering
\includegraphics[width=1.0\textwidth, keepaspectratio]{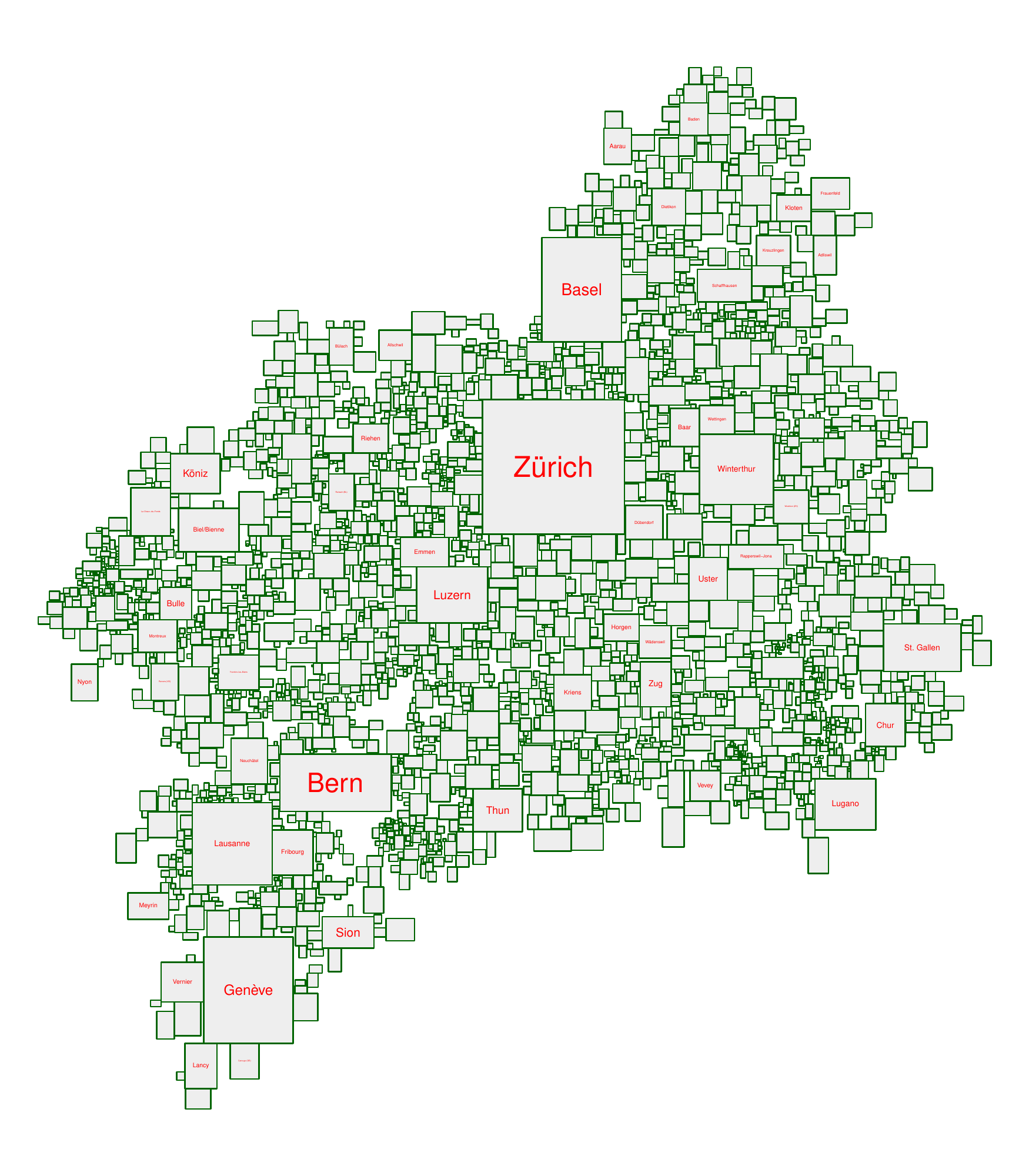}
\caption{A rectangular population cartogram of Switzerland is shown.
map data source: Swiss Federal Office of Topography using
\href{http://www.toposhop.admin.ch/en/shop/products/landscape/gg25_1}{Landscape Models / Boundaries GG25}, downloaded 2016-05-01;
statistical data:
\href{http://www.bfs.admin.ch/bfs/portal/de/index.html}{Bundesamt f{\"u}r Statistik (BFS), Website Statistik Schweiz},
downloaded file
\href{http://www.bfs.admin.ch/bfs/portal/de/index/regionen/02/daten.html}{je-d-21.03.01.xls}
on 2016-05-26.
\label{figure:Switzerland:population}
An interactive shiny app is available through \url{https://recmap.shinyapps.io/CH-Gemeinden/}. }
\end{figure}

\paragraph{A population cartogram of Switzerland} on
community (Gemeinde) level is drawn in
Figure~\ref{figure:Switzerland:population}. The
rectangles of the original map were extracted from an
ESRI shape file of the map data Landschaftsmodelle: GG25
from the Federal Office of Topography (swisstopo) using
\pkg{shapefiles} by~\cite{shapefiles}. The following
attributes were extracted for each map region:
\code{box}, \code{Gemeindecode}, and \code{Gemeindename}.
There are 2300 rectangles to place. The statistical
values (population 2013, published in 2015) were
downloaded from Swiss Statistics (Regionalportr{\"a}ts:
Kennzahlen aller Gemeinden (je-d-21.03.01) Bundesamt
f{\"u}r Statistik BFS
\url{http://www.bfs.admin.ch/bfs/portal/de/index/regionen/02/daten.html})
and joined by the \code{Gemeindecode} attribute with the
swisstopo map. On the web application, available through
\url{https://recmap.shinyapps.io/CH-Gemeinden/}, the
reader can play with the rectangular population cartogram
of Switzerland and different statisical values provided
in the previous mentioned data for coloring the cartogram
regions.

\begin{figure}[htbp]
\centering
\includegraphics[width=1.0\textwidth, keepaspectratio]{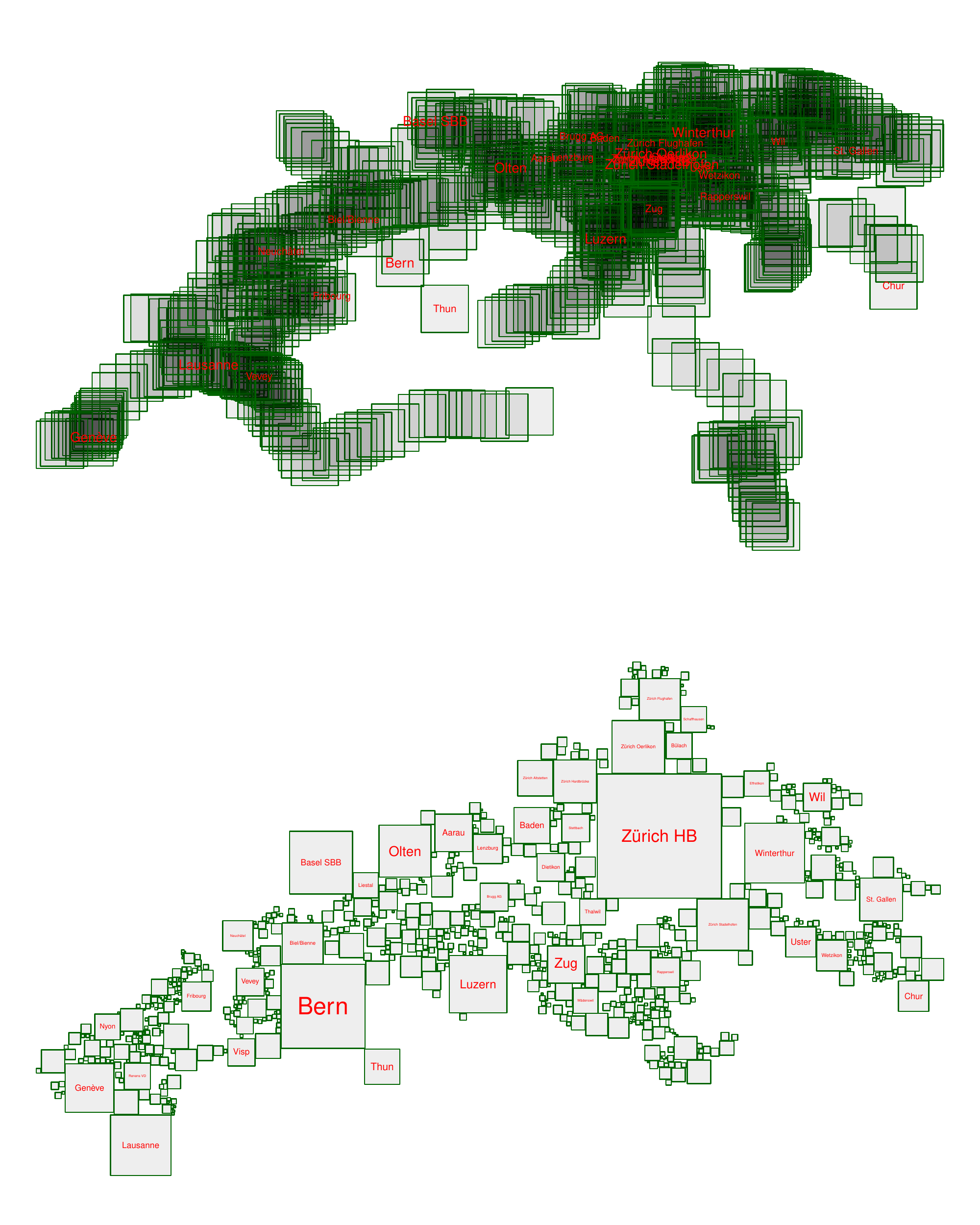}
\caption{A Swiss railway passenger frequency cartogram is shown on the lower map. The upper graphic displays the overlapping rectangles of the input map. 
source: \url{sbb.ch}, 2016-05-12.}
\label{figure:sbb} 
\end{figure}

\begin{figure}[htbp]
\centering
\includegraphics[height=0.25\textwidth, keepaspectratio]{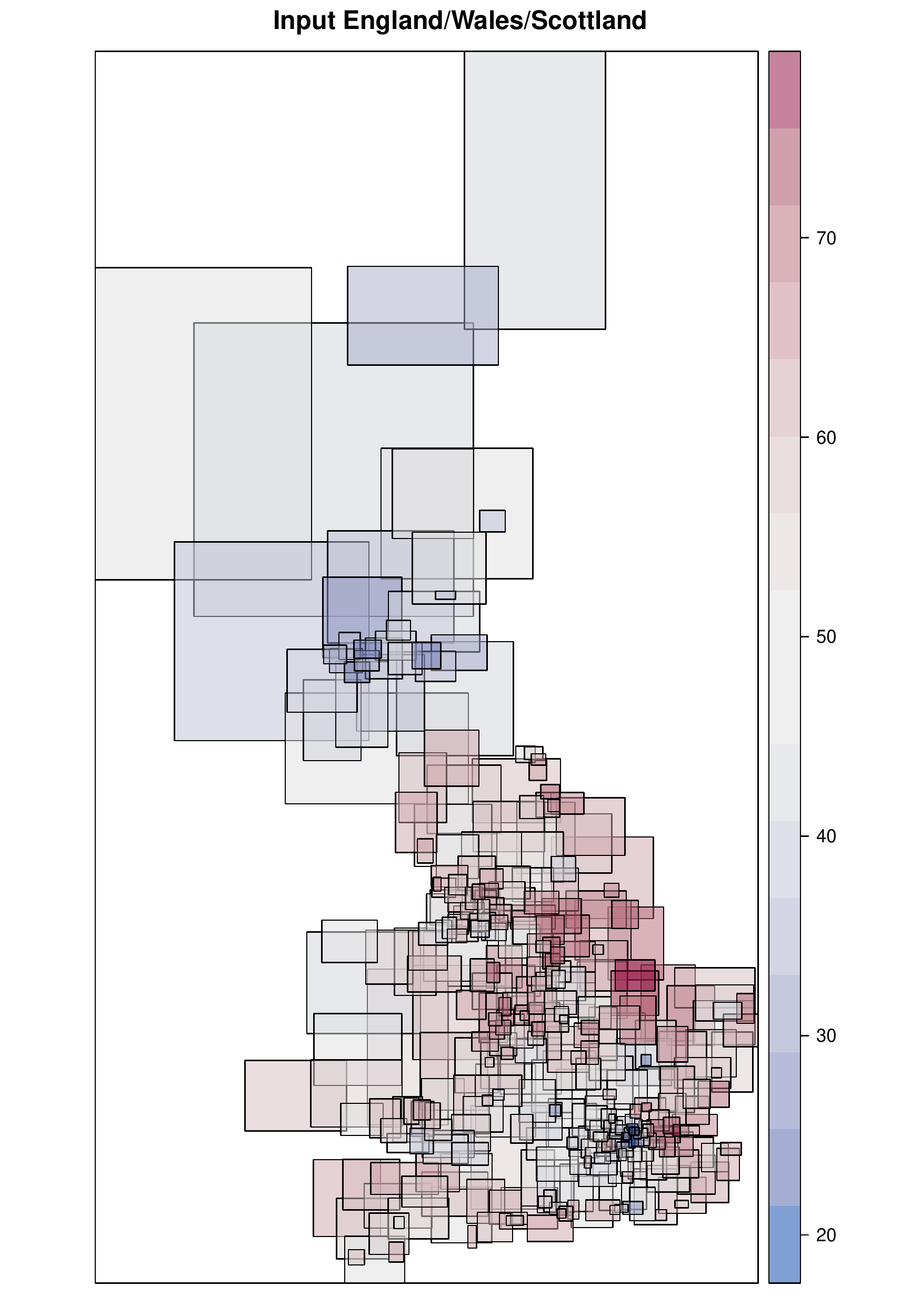}
\includegraphics[height=0.25\textwidth, keepaspectratio]{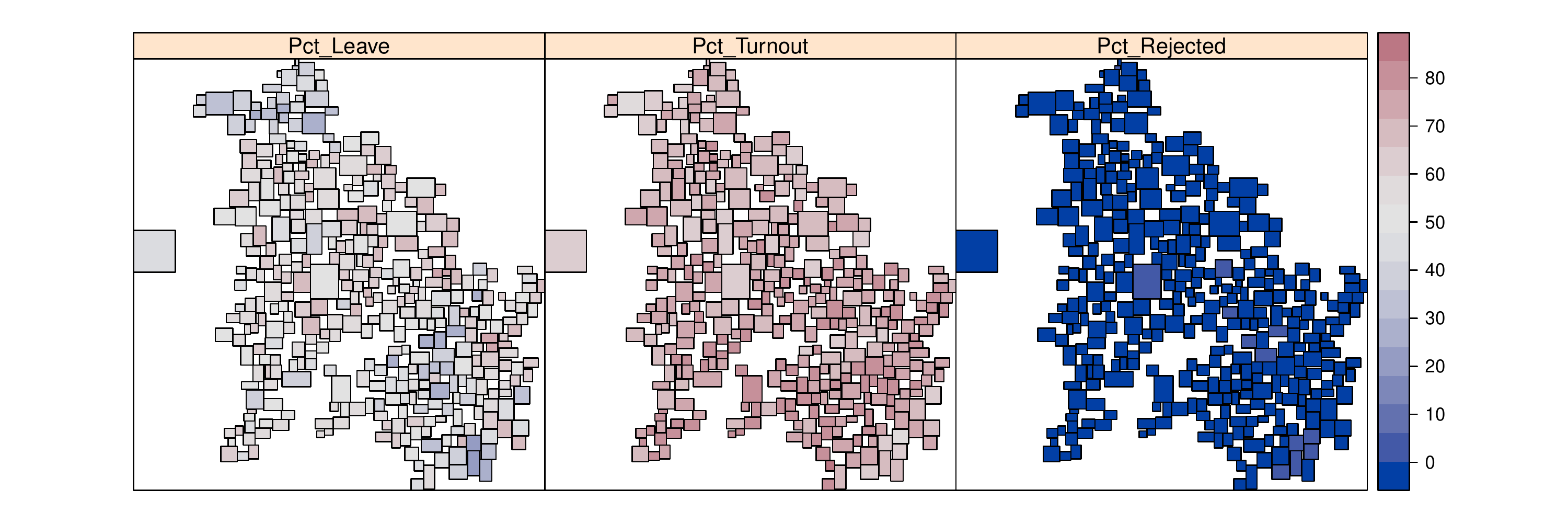}
\includegraphics[width=0.8\textwidth, keepaspectratio]{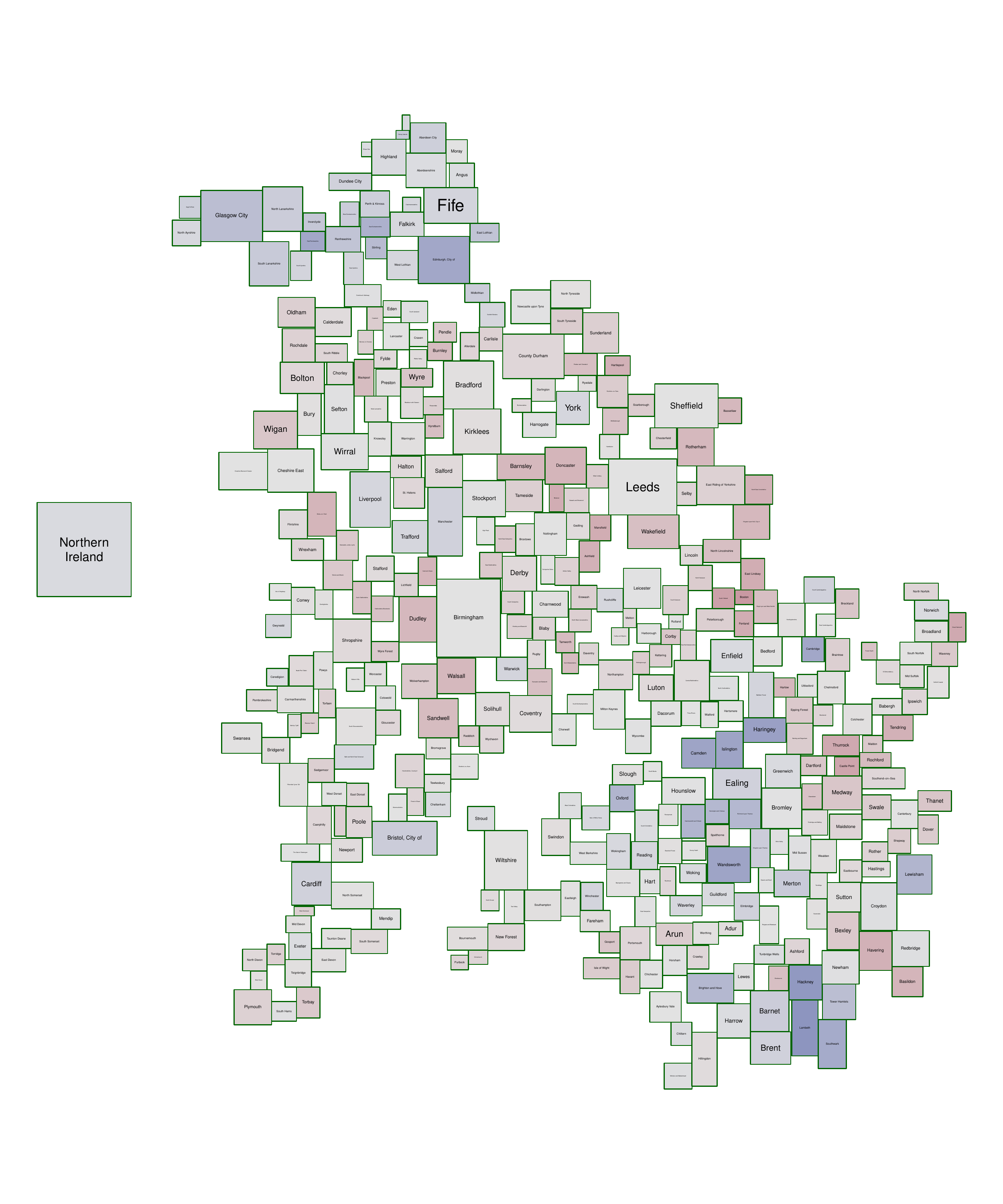}
\caption{The outcome of UK Brexit EU-referendum is displayed.
Northern Ireland was manually added. The overlapping MBBs of the input map
are displayed in the top left.
On all other plots, the region areas represent the
number of the electorates. The colors are indicating the outcome of the referendum
(blue: remain / red: leave;
as lower the color intensity as closer to 50\% : 50\% outcome ).
Other attributes represented as percentages (Pct) are displayed using the
\code{spplot} of the \pkg{sp} package (top right).
\label{fig:brexit}
Copyright: Contains National Statistics data~\copyright~Crown copyright and database right 2016
Contains NRS data~\copyright~Crown copyright and database right 2016
Source: NISRA : Website: \url{www.nisra.gov.uk}
Contains OS data~\copyright~Crown copyright [and database right] (2016)
}
\end{figure}

\paragraph{A Swiss railway passenger frequency cartogram}
is graphed on the bottom of Figure~\ref{figure:sbb}.
The upper visualization shows the overlapping rectanglues
of all 400 geo-locations which define the pseudo dual of the map.
The data were retrieved from
\url{https://data.sbb.ch/explore/}
and contain already the longitude and latitude coordinates of of the
railway main station and stops.

\paragraph{The UK Brexit EU-referendum} is shown as final example in
Figure~\ref{fig:brexit}.
The UK boundary file were downloaded from \url{https://census.edina.ac.uk}
and joined by the column name \code{geo_code} and \code{Area_Code} with the
the outcome of the referendum downloaded through
\url{http://www.electoralcommission.org.uk/} on July 3th.
This example also demonstrate the usage of the \pkg{sp} package
by~\cite{sp2013}.
Through using the function \code{recmap2sp} the recmap class can be transformed
into a \code{SpatialPolygonsDataFrame} object. The following code snippets
show how the \code{summary} and \code{spplot} methods can be used.

\begin{Schunk}
\begin{Sinput}
R> DF <- rbind(data.frame(Pct_Leave = UK$Map$Pct_Leave,
+                   Pct_Turnout = UK$Map$Pct_Turnout,
+                   Pct_Rejected = UK$Map$Pct_Rejected,
+                   row.names = UK$Map$name),
+              data.frame(Pct_Leave =44.22,
+              Pct_Turnout = 62.69,
+              Pct_Rejected = 0.05,
+              row.names = 'Northern\nIreland'))
R> UK.sp <- recmap2sp(add_NI(UK.recmap) , DF)
R> summary(UK.sp)
\end{Sinput}
\begin{Soutput}
Object of class SpatialPolygonsDataFrame
Coordinates:
        min     max
x -121259.4 1389447
y -340948.2 1243144
Is projected: NA 
proj4string : [NA]
Data attributes:
   Pct_Leave      Pct_Turnout     Pct_Rejected    
 Min.   :21.38   Min.   :56.25   Min.   :0.03000  
 1st Qu.:47.38   1st Qu.:70.19   1st Qu.:0.06000  
 Median :54.34   Median :74.30   Median :0.07000  
 Mean   :53.29   Mean   :73.69   Mean   :0.07283  
 3rd Qu.:60.49   3rd Qu.:77.89   3rd Qu.:0.08000  
 Max.   :75.56   Max.   :83.57   Max.   :0.24000  
\end{Soutput}
\begin{Sinput}
R> spplot(UK.sp, col.regions=diverge_hcl(19)[1:16], layout=c(3,1))
\end{Sinput}
\end{Schunk}

\begin{sidewaystable}
\resizebox{\textwidth}{!}{
\begin{tabular}{ll|rr|lrrr|lr|rl|rr}
\hline
\hline
map	&	statistical value	&	\#regions	&	area err	&	meta&	pop size	&	mutrate	&	\#gen	& compute hardware     	&	cores	&	proc time	&	unit	&	&	Fig\\
\hline
US state level &  \#electors 	&	49	&	0.36	&	GA	&	unknown	&	unknown	&	10	&	Intel Xeon CPU @ 1.5 GHz	&	1	&	$\approx$1	&	min	&		&	\ref{figure:USelection}\\
US state level &  area 	&	50	&	0.17	&	-	&	-&	-&	-	&	-	&	1	&	$\ll$1	&	sec	&		&	\ref{figure:usage}\\
8x8 checkerborad	&	1:4	&	64	&	0.27	&	GA	&	256	&	0.25	&     533  	&	3 GHz Intel Core i7	&	4	&	$\approx$1	&	min   &		&	\ref{figure:cmp_GA_GRASP}\\
8x8 checkerborad	&	1:4	&	64	&	0.27	&	GRASP	&	NA	&	NA	&	NA	&	3 GHz Intel Core i7	&	4	&	$\approx$1	&	min	&		&	\ref{figure:cmp_GA_GRASP}\\
\hline
US state level	&	population 1977	&	48	&      0.41	&	GA	&	300	&	0.25	&	11	&	3 GHz Intel Core i7	&	4	&	5	&	sec	&		&	\ref{figure:x77}\\
US state level	&	murder 1977	&	48	&      0.37	& 	GA	&	300	&	0.25	&	20	&	3 GHz Intel Core i7	&	4	&	10	&	sec	&		&	\ref{figure:x77}\\
US state level	&	income 1977	&	48	&      0.29 	&	GA	&	300	&	0.25	&	30	&	3 GHz Intel Core i7	&	4	&	16	&	sec	&		&	\ref{figure:x77}\\
US state level	&	illiteracy 1977     &	48	&      0.40 	&	GA	&	300	&	0.25	&	30	&	3 GHz Intel Core i7	&	4	&	16	&	sec	&		&	\ref{figure:x77}\\
\hline
US California	&	population 2010	&	48	&	0.62	&	GA	&	240	&	0.25	&	27	&	3 GHz Intel Core i7	&	4	&	10	&	sec	&		&	\ref{figure:us_census}\\
US Colorado	&	population 2010	&	64	&	0.68	&	GA	&	320	&	0.25	&	102	&	3 GHz Intel Core i7	&	4	&	74	&	sec	&		&	\ref{figure:us_census}\\
US Florida	&	population 2010	&	68	&	0.44	&	GA	&	240	&	0.25	&	49	&	3 GHz Intel Core i7	&	4	&	39	&	sec	&		&	\ref{figure:us_census}\\
US New Jersey	&	population 2010	&	21	&	0.38	&	GA	&	105	&	0.25	&	30	&	3 GHz Intel Core i7	&	4	&	4	&	sec	&		&	\ref{figure:us_census}\\
US New York	&	population 2010	&	62	&	0.62      &	GA	&	310	&	0.25	&	148	&	3 GHz Intel Core i7	&	4	&	106	&	sec	&		&	\ref{figure:us_census}\\
\hline
Switzerland	&	population 2010	&	2300	&	0.59	&	GA	&	1280	&	0.35	&	300	& 	Intel Xeon CPU E5-2698 v3 @ 2.30GHz	&	64	&	3	&	days	&		&	\ref{figure:Switzerland:population}\\
Swiss SBB railway	&	passenger frequency	&	724	&	NA	&      GA	&	 1000	&	0.25	&  1000     	&	Intel Xeon CPU E5-2698 v3 @ 2.30GHz	&	64	&	2	&	days	&		&	\ref{figure:sbb}\\
UK	&	electorates	&	370	&	0.57	&	GA	&	1200	&	0.25	&	4000	&	Intel Xeon CPU E5-2698 v3 @ 2.30GHz	&	64	&	3	&	days	&		&	\ref{fig:brexit}\\
\hline
\hline
\end{tabular}
}
\caption{
The spreadsheet provides an overview of all statistical rectangular cartograms displayed in this manuscript ordered by appearance. The spreadsheet includes input properties, e.g., number of map regions, as well as process time on the used hardware platforms. The area error was computed using the \code{summary.recmap} method implementing the area error function introduced by \cite{cartodraw}.
Since the Swiss railway passenger frequency cartogram is derived from a 2D point set, it is not meaningful to compute an area error of the input.
The rectangular cartograms having more than 300 map regions required more generation of the genetic algorithm and were computed on a compute server due to the higher computing demands.
}
\label{table:overview}
\end{sidewaystable}
\section{Summary}
\label{section:summary}

This article introduces the CRAN \pkg{recmap} package which
implements the RecMap MP2 algorithm. This method generates
rectangular statistical cartograms. Two outstanding
features of the implemented algorithm are: the areas of the
map regions represent the exact statistical value without
any area error and the ratios of the map regions are
preserved. These constraints are important for the correct
interpretation of the geography-related statistical data.
It is evident that using these restrictions the map
topology can not be preserved. The implementation allows
the generation of rectangular statistical cartograms having
less than one hundred map regions within a few seconds with
support of a metaheuristic. The cartograms generated enable
an interactive explorative data analysis. All necessary
steps can be done on the R command line or by using web
applications on a modern computer. It has been
demonstrated, how the drawing of the cartogram can be
optimized according to a fitness function by using a
metaheuristic and benefiting from today's multi core
hardware and R's parallel environment. Most promising is
using a fitness function which is derived from the relative
position error objective function. It should also be
highlighted that the method can read a spreadsheet
containing the geographic location. It does not require any
complex polygon mesh as input.
The potential of the method is shown by using
real world maps covering a maps size of three orders
magnitude and synthetical data (8x8 checkerboard). 
Table~\ref{table:overview} provides an overview of the 
rectangular cartogram specification.
The \pkg{recmap} package is a powerful tool in the hand of data
analysts, cartographers, or statisticians using R who want
to draw their own statistical rectangular cartograms.
\bibliography{recmap}
\end{document}